\documentclass[twocolumn, tighten]{aastex63}
\usepackage{natbib}
\usepackage{needspace}
\usepackage{tabularx}
\usepackage{booktabs}
\usepackage{multirow}
\usepackage{amsmath}
\usepackage{tabularx} 
\usepackage{tablefootnote}
\let\tablenum\relax % Let siunitx redefine tablenum
\usepackage{siunitx}
\newcommand{\tess}{\textit{TESS}}

\newcommand{\ktwo}{{\textit K2}}
\newcommand{\Gaia}{{\textit{Gaia} DR3}}
\newcommand\vsini{$v\sin{i}$}
\newcommand\Msun{$M_{\odot}$} %
\newcommand\Rsun{$R_{\odot}$} %
\newcommand\Lsun{$L_{\odot}$} %
\newcommand\RE{$R_{\Earth}$} % 

\newcommand{\obsstart}{2019 December 24}
\newcommand{\obsend}{2022 January 28}
\newcommand{\name}{HD~63433}
\newcommand{\pname}{\texorpdfstring{HD~63433\,d}{HD~63433~d}} % hyperref complains that the PDF bookmark cannot contain the "\," space. The second argument to \texorpdfstring is the alternative text that latex should write to the PDF bookmark
\newcommand{\mass}{$0.99 \pm 0.03$} %Mann+2020
\newcommand{\age}{$414 \pm 23$} %Jones+2015
\newcommand{\gaiamag}{$6.7372 \pm 0.0028$} %gaia dr3 
\newcommand{\dist}{$22.32 \pm 0.06$} %gaia dr3
\newcommand{\rad}{$0.912 \pm 0.034$} %mann+2020
\newcommand{\prad}{$1.073^{ +0.046}_{-0.044}$\,\RE{}}
\newcommand{\porb}{$4.209075^{ +0.000012}_{-0.000023}$\,days}

\newcommand{\semimajoraxis}{$0.0503^{ +0.0025}_{-0.0027}$\,AU}
\newcommand{\feh}{$0.03 \pm 0.05$} %Hinkel
\newcommand{\minmass}{0.2} 
\newcommand{\maxmass}{1.6} 

\shortauthors{Capistrant, Soares-Furtado et al.}
\begin{document}
\title{TESS Hunt for Young and Maturing Exoplanets (THYME) XI: An
Earth-sized Planet Orbiting a Nearby, Solar-like Host in the 400\,Myr Ursa Major Moving Group}
\author[0000-0002-4592-8799]{Benjamin K.~Capistrant}
\email{bcapistrant@ufl.edu}
\affiliation{Department of Astronomy, University of Florida, Bryant Space Science Center, Stadium Road, Gainesville, FL 32611, USA }

\author[0000-0001-7493-7419]{Melinda Soares-Furtado}
\email{mmsoares@wisc.edu}
\altaffiliation{Joint first author}
\altaffiliation{NASA Hubble Postdoctoral Fellow}
\affiliation{Department of Astronomy,  University of Wisconsin-Madison, 475 N.~Charter St., Madison, WI 53706, USA}
\affiliation{Department of Physics and Kavli Institute for Astrophysics and Space Research, Massachusetts Institute of Technology, Cambridge, MA 02139, USA}

\author[0000-0001-7246-5438]{Andrew Vanderburg}
\affiliation{Department of Physics and Kavli Institute for Astrophysics and Space Research, Massachusetts Institute of Technology, Cambridge, MA 02139, USA}

\author[0000-0003-4287-004X]{Alyssa Jankowski}
\affiliation{Department of Astronomy,  University of Wisconsin-Madison, 475 N.~Charter St., Madison, WI 53706, USA}

\author[0000-0003-3654-1602]{Andrew W.~Mann}
\affiliation{Department of Physics \& Astronomy, The University of North Carolina at Chapel Hill, Chapel Hill, NC 27599-3255, USA}

\author[0009-0006-7023-1199]{Gabrielle Ross}%
\affiliation{Department of Astrophysical Sciences, Princeton University, Princeton, NJ 08544, USA}

\author{Gregor Srdoc}
\affiliation{Kotizarovci Observatory, Sarsoni 90, 51216 Viskovo, Croatia}

\author[0000-0003-0595-5132]{Natalie R.~Hinkel}
\affiliation{Physics \& Astronomy Department, Louisiana State University, Baton Rouge, LA 70803, USA}
\affiliation{Southwest Research Institute, San Antonio, TX 78238, USA}

\author[0000-0002-7733-4522]{Juliette Becker}
\affiliation{Department of Astronomy,  University of Wisconsin-Madison, 475 N.~Charter St., Madison, WI 53706, USA}

\author[0000-0001-6343-4744]{Christian Magliano}
\affiliation{Dipartimento di Fisica ``Ettore Pancini'', Università di Napoli Federico II, 80126 Napoli, Italy}

\author[0000-0002-9521-9798]{Mary Anne Limbach}
\affiliation{Department of Astronomy, University of Michigan, Ann Arbor, MI 48109, USA}

\author[0000-0001-8220-0548]{Alexander P. Stephan}
\affiliation{Department of Astronomy, The Ohio State University, Columbus, OH 43210, USA}
\affiliation{Center for Cosmology and AstroParticle Physics, The Ohio State University, Columbus, OH 43210, USA}

\author[0000-0002-6478-0611]{Andrew C.~Nine}
\affiliation{Department of Astronomy,  University of Wisconsin-Madison, 475 N.~Charter St., Madison, WI 53706, USA}

\author[0000-0003-2053-0749]{Benjamin M.~Tofflemire}
\altaffiliation{51 Pegasi b Fellow}
\affiliation{Department of Astronomy, The University of Texas at Austin, Austin, TX 78712, USA}

\author[0000-0001-9811-568X]{Adam L.~Kraus}
\affiliation{Department of Astronomy, The University of Texas at Austin, Austin, TX 78712, USA}

\author[0000-0002-8965-3969]{Steven Giacalone}
\affiliation{Department of Astronomy, 501 Campbell Hall \#3411, University of California, Berkeley, Berkeley, CA 94720, USA}

\author[0000-0002-4265-047X]{Joshua N.~Winn}
\affiliation{Department of Astrophysical Sciences, Princeton University, Princeton, NJ 08544, USA}

\author[0000-0001-6637-5401]{Allyson Bieryla}
\affiliation{Center for Astrophysics ${\rm \mid}$ Harvard {\rm \&} Smithsonian, 60 Garden Street, Cambridge, MA 02138, USA}

\author[0000-0002-0514-5538]{Luke~G.~Bouma}
\altaffiliation{51 Pegasi b Fellow}
\affiliation{Department of Astronomy, MC 249-17, California Institute of Technology, Pasadena, CA 91125, USA}

\author[0000-0002-5741-3047]{David~R.~Ciardi}
\affiliation{Caltech/IPAC-NASA Exoplanet Science Institute, 770 S. Wilson Avenue, Pasadena, CA 91106, USA}

\author[0000-0001-6588-9574]{Karen~A.~Collins}
\affiliation{Center for Astrophysics ${\rm \mid}$ Harvard {\rm \&} Smithsonian, 60 Garden Street, Cambridge, MA 02138, USA}

\author[0000-0002-2553-096X]{Giovanni Covone}
\affiliation{Dipartimento di Fisica ``Ettore Pancini'', Università di Napoli Federico II, 80126 Napoli, Italy}

\author[0000-0002-7564-6047]{Zo{\"e} L.~de Beurs}
\altaffiliation{NSF Graduate Research Fellow}
\altaffiliation{MIT Presidential Fellow}
\affiliation{Department of Earth, Atmospheric and Planetary Sciences, Massachusetts Institute of Technology, Cambridge, MA 02139, USA}

\author[0000-0003-0918-7484]{Chelsea~X.~Huang}
\affiliation{University of Southern Queensland, Centre for Astrophysics, West Street, Toowoomba, QLD 4350 Australia}

\author[0000-0002-4715-9460]{Jon M.~Jenkins}
\affiliation{NASA Ames Research Center, Moffett Field, CA 94035, USA}

\author[0000-0003-0514-1147]{Laura Kreidberg}
\affiliation{Max Planck Institute for Astronomy, K\"{o}nigstuhl 17, D-69117 Heidelberg, Germany}

\author[0000-0001-9911-7388]{David W.~Latham}
\affiliation{Center for Astrophysics ${\rm \mid}$ Harvard {\rm \&} Smithsonian, 60 Garden Street, Cambridge, MA 02138, USA}

\author[0000-0002-8964-8377]{Samuel~N.~Quinn}
\affiliation{Center for Astrophysics ${\rm \mid}$ Harvard {\rm \&} Smithsonian, 60 Garden Street, Cambridge, MA 02138, USA}

% Architect
\author[0000-0002-6892-6948]{Sara Seager}
\affiliation{Department of Physics and Kavli Institute for Astrophysics and Space Research, Massachusetts Institute of Technology, Cambridge, MA 02139, USA}
\affiliation{Department of Earth, Atmospheric and Planetary Sciences, Massachusetts Institute of Technology, Cambridge, MA 02139, USA}
\affiliation{Department of Aeronautics and Astronautics, MIT, 77 Massachusetts Avenue, Cambridge, MA 02139, USA}

\author[0000-0002-1836-3120]{Avi Shporer}
\affiliation{Department of Physics and Kavli Institute for Astrophysics and Space Research, Massachusetts Institute of Technology, Cambridge, MA 02139, USA}

\author[0000-0002-6778-7552]{Joseph~D.~Twicken}
%\affiliation{SETI Institute, Moffett Field, CA, 94035, USA}
\affiliation{SETI Institute, Mountain View, CA 94043 USA/NASA Ames Research Center, Moffett Field, CA 94035 USA}

\author[0000-0002-5402-9613]{Bill~Wohler}
\affiliation{SETI Institute, Mountain View, CA 94043 USA/NASA Ames Research Center, Moffett Field, CA 94035 USA}

\author[0000-0001-6763-6562]{Roland K.~Vanderspek}
\affiliation{Department of Physics and Kavli Institute for Astrophysics and Space Research, Massachusetts Institute of Technology, Cambridge, MA 02139, USA}

\author[0000-0003-0381-1039]{Ricardo Yarza}
\altaffiliation{NASA FINESST Fellow}
\altaffiliation{Frontera Computational Science Fellow}
\affiliation{Department of Astronomy and Astrophysics, University of California, Santa Cruz, CA 95064, USA}
\affiliation{Texas Advanced Computing Center, University of Texas, Austin, TX 78759, USA}

\author[0000-0002-0619-7639]{Carl Ziegler}
\affiliation{Department of Physics, Engineering and Astronomy, Stephen F.~Austin State University, 1936 North St, Nacogdoches, TX 75962, USA}

\begin{abstract}
Young terrestrial worlds are critical test beds to constrain prevailing theories of planetary formation and evolution. 
We present the discovery of \pname{} --- a nearby (22\,pc), Earth-sized planet transiting a young sunlike star (TOI-1726, \name{}). 
\pname{} is the third planet detected in this multiplanet system. 
The kinematic, rotational, and abundance properties of the host star indicate that it belongs to the young (\age{}\,Myr) Ursa Major moving group, whose membership we update using new data from \Gaia{} and TESS.
Our transit analysis of the \tess{} light curves indicates that \pname{} has a radius of 1.1\,\RE{} and closely orbits its host star with a period of 4.2\,days.
To date, \pname{} is the smallest confirmed exoplanet with an age less than $500$\,Myr, 
%The system's proximity makes this new detection the nearest young Earth-sized world yet observed. 
and the nearest young Earth-sized planet. 
Furthermore, the apparent brightness of the stellar host ($V\simeq6.9$\,mag) makes this transiting multiplanet system favorable to further investigations, including spectroscopic follow-up to probe atmospheric loss in a young Earth-sized world. 
\end{abstract}


\section{Introduction}
\label{sec:intro}

The first few hundred million years after formation is a critical stage of the planetary lifecycle. At this early evolutionary stage, planets may thermally contract \citep[e.g.,][]{Fortney2011} and  experience  significant atmospheric losses \citep[e.g.,][]{Lammer2003, Lopez2013,Ginzburg2018}.
Moreover, the early phase of planetary system formation can be very dynamically active, particularly after the dissipation of the proto-planetary gas disk \citep[e.g.,][]{1998Icar..136..304C,2012AREPS..40..251M}.
During the first tens of millions of years, young systems may be susceptible to planet-planet scattering events \citep[e.g.,][]{Fabrycky2007,Chatterjee+2008,Ford+2008,Naoz2011}, planet collisions \citep[e.g.,][]{Asphaug+2006,Hansen2009}, and the prolonged bombardment of planets by asteroids and planetesimals. Additionally, planetary systems in dense star-forming regions may undergo dynamical interactions with other stars in their natal clusters \citep[e.g.,][]{Fregeau+2006}. 
%, further shaping the dynamical history of young planets for the lifetime of these stellar formation regions
Detailed observations of planetary systems in such environments are, therefore, crucial to understanding the general formation history of the exoplanet population.

%MSF to double-check the number below before final submission
Among the population of confirmed exoplanets with precisely-measured age estimates, 50 exoplanets are estimated to be younger than 500\,Myr.
For context, we considered a system age estimate to be precisely measured if its error estimate is at most 50\% of the estimated age value.
Rarer still are young multiplanet systems with hosts bright enough to permit follow-up precision radial velocity investigations --- only nine young multiplanet hosts are brighter than $G=12$\,mag.

In this paper, we report the detection and characterization of \pname{} --- a young (\age{}\,Myr), nearby (\dist{}\,pc) Earth-sized planet orbiting a sunlike host (spectral type G5V). 
\pname{} is the third planet to be detected in this multiplanet system. 
The discovery of the other two \name{} planets, both mini-Neptunes, was reported by \cite{Mann2020}. 

The discovery of young transiting exoplanets that are amenable to follow-up investigations is of critical importance, as such observations offer constraints to prevailing theoretical models of planetary formation and evolution. 
Because the planets in the \name{} system transit a bright ($V\simeq6.9$\,mag) host star, this system offers a good opportunity for follow-up analyses.
This includes investigations of the transmission spectrum, stellar obliquity, and mass estimates derived via ground-based radial velocity data. The \name{} system has already been the subject of numerous follow-up studies thanks to its brightness and proximity \citep{dai, zhang2022, damasso, mallorquin}, and the newly detected \pname{} provides an additional opportunity for follow-up investigations, with the potential to offer constraints on the process of atmospheric escape in a young, Earth-sized exoplanet. 

Our analysis of \pname{} is part of the \tess{} Hunt for Young and Maturing Exoplanets (THYME; \citealt{Newton2019}). 
The principal goal of THYME is the identification of young transiting exoplanets in moving groups, stellar associations, and open clusters.
Stellar groups play an important role in this search, as they are comprised of stars that share common ages and compositions.
Therefore, the ages of these systems may be determined using a combination of methods.
The THYME project leverages the \textit{Transiting Exoplanet Survey Satellite} (\tess{}; \citealt{Ricker2015}) light curves\footnote{In this paper, we use the terms ``time series data" and ``light curves" interchangeably.} of stellar members to search for the periodic brightness changes that are indicative of transiting exoplanets.
For a review of the transit detection technique, we refer the reader to \cite{Santos2008}.
The THYME project is the successor to the Zodiacal Exoplanets in Time Survey (ZEIT; \citealt{Mann2016}), which performed an analogous transiting exoplanet search using time series data from the \ktwo{} mission.

The following outline presents the organization of this manuscript. 
In Section~\ref{sec:data}, we present the data used in our analysis of \name{} and other members of the Ursa Major moving group (hereafter referred to as ``UMa").
In Section~\ref{sec:properties}, we present the observed and inferred properties of \name{}. 
In Section~\ref{sec:analysis}, we present our analysis of the transit signatures found in the \tess{} light curve of the host star.
In this section, we also present our UMa membership analysis for the planet host, as well as the broader sample of likely UMa members.
In Section~\ref{sec:Vetting}, we describe the procedures we performed to assess false positive scenarios for the \pname{} candidate.
In Section~\ref{sec:results}, we present the results we obtained in our exoplanet characterization, as well as the results from our photometric rotational analyses.
Finally, in Section~\ref{sec:summary}, we present our summary and conclusions, highlighting some potential follow-up work.

\section{Data}
\label{sec:data}
\subsection{TESS Photometry}
\subsubsection{The HD 63433 TESS Light Curve}
\label{subsubsec:planetsearch}
\tess{} observations are divided into individual sectors.
Each sector is observed continuously for approximately 27 days and covers a region $24^{\circ} \times 96^{\circ}$ in size. 
To date, \name{} has been observed in five distinct \tess{} sectors, resulting in an irregularly sampled baseline that spans between \obsstart{} to \obsend{}. 
In Table~\ref{tab:sectors} we list the sectors, camera numbers, CCD numbers, and sector-specific date ranges that correspond to each observation. 

\begin{table}[ht!]
\centering
\begin{tabular}{|l|l|l|l|l|l}
\cline{1-5}
Sector & Cadence & Camera & CCD & Date Range \\ \cline{1-5}
20 & 2\,min &1 & 4 & 2019/12/24 -- \\
 & & &  & 2020/01/20\\ \cline{1-5}
44 & 20\,sec &4 & 1 & 2021/10/12 -- \\
 & & &  &  2021/11/06\\ \cline{1-5}
45 & 20\,sec &2 & 3 & 2021/11/06	--  \\ 
 & & &  & 2021/12/02\\ \cline{1-5}
46 & 20\,sec &1 & 3 & 2021/12/02	-- \\
 & & &  & 2022/12/30\\ \cline{1-5}
47 & 20\,sec &1 & 4 & 2021/12/30 -- \\
 & & &  & 2022/01/28\\ \cline{1-5}
\end{tabular}
\caption{The observation specifications for each of the \tess{} sectors analyzed in the detection and characterization of \pname{}.}
\label{tab:sectors}
\end{table}

\vspace{-6.0mm}
To analyze the \name{} time series and search for exoplanetary signals, we leveraged data from all five \tess{} sectors; the image data were reduced and analyzed by the Science Processing Operations Center at NASA Ames Research Center \citep{jenkinsSPOC2016}.
It has been shown that 20-second cadence data has better photometric precision than 2-minute data \citep{Huber2022}. 
Therefore, we used 20-second cadence data when it was available, opting for 2-minute cadence data otherwise. 
Also, in our experience, a custom systematics correction can often yield improved results over the default \tess{} pipeline light curves, especially for young stars with high amplitude stellar variability. We performed a bespoke analysis, starting with the simple aperture photometry (SAP; \citealt{Twicken2010SAP}; \citealt{Morris2020SAP}) light curves generated by the \tess{} Science Processing Operations Center (SPOC) pipeline.\footnote{SPOC pipeline description: \url{https://heasarc.gsfc.nasa.gov/docs/tess/pipeline.html}} 
The SPOC data described in this section can be found in MAST: \dataset[10.17909/t51r-sh87]{http://dx.doi.org/10.17909/t51r-sh87}.

We applied systematic corrections to these light curves, following the procedure outlined by \cite{Vanderburg2019} and briefly summarized here.
The SPOC light curves were fit with a linear model, which consisted of the following components:
\begin{itemize}
    \item A basis spline (B-spline) with regularly-spaced breaks at 0.3-day intervals to model signatures of low-frequency stellar variability.
    \item The mean and standard deviation of the spacecraft quaternion time series within each light curve exposure.\footnote{The quaternion measurements are two-second-cadence vector time series that describe the spacecraft orientation in relation to guide star observations. \tess{} quaternions are available at \url{https://archive.stsci.edu/missions/tess/engineering/}.}
    \item Seven co-trending vectors from the SPOC Presearch Data Conditioning’s band 3 (fast timescale) flux time series correction with the largest eigenvalues \citep{Stumpe2012,Stumpe2014,smith2012}.
    \item A high-pass-filtered (0.1\,days) flux time series from the SPOC background aperture.
\end{itemize} 

\noindent No flares were observed in the data. Leveraging the systematics-corrected light curves for all five \tess{} sectors, we searched for transit signals using a Box-Least-Squares (BLS) periodogram \citep{Kovacs2002}. 
This methodology is described in more detail in \cite{Vanderburg2016}.
We detected the transiting signatures of exoplanets \name{}\,b and \name{}\,c at high significance. Both HD 63433 b and c were detected by the Science Processing Operations Center \citep{jenkinsSPOC2016} at NASA Ames Research Center on 2 February 2020, and the TESS Science Office issued alerts for these candidates on 19 February 2020 and 20 February 2020, respectively \citep{Guerrero2021}.
\cite{Mann2020} first reported the discovery of these planets using a single sector of \tess{} data (Sector 20). 
After removal of those two signals, we detected for the first time a third transit signature with a period of 4\,days and a shallower transit depth than the signatures of \name{}\,b and \name{}\,c. {We additionally apply this process using the Transit least squares (TLS) algorithm \citep{HippkeTLS2019}, which incorporates features such as stellar limb-darkening and the effects of planetary ingress and egress. We recover the third transit signature using this algorithm, and as found with our BLS search, see no additional periodic signals after removing all three transits from the light curve.}

We attribute this signal to a new planet called \pname{}. The transit depth of \pname{} is about 100\,ppm, consistent with that of an Earth-sized planet. The new signal was not reported by either the \tess{} SPOC or MIT Quick Look Pipeline (QLP; \citealt{QLP, QLP2}) transit searches and was not listed as a \tess{} planet candidate. 
We found that the \pname{} transit signal was too shallow to be detected with a single sector of \tess{} data, which is likely why it escaped detection in \cite{Mann2020}.
Nevertheless, this transit signal is clearly detectable in our analysis of the full, five-sector light curve, with a signal-to-noise ratio of about 15. % We hypothesize that stellar variability precluded the detection of this signal by either SPOC or QLP in their searches of the full five-sector light curve. 

We performed the standard battery of tests to assess whether the origin of the signal is indeed planetary in nature.
This includes an inspection of the phase-folded light curve, tests for consistency of even/odd transits, and a search for secondary eclipses. 
We found no evidence to refute the detection of \pname{} and proceed under the assumption that this signal is a viable planet candidate. 
We also performed a difference image centroid offset analysis \citep{Twicken2018} using the SPOC data validation pipeline software and found that the signal must originate within about 20\,arcseconds of \name{}, corroborating this interpretation.  

After producing our final light curve, we flattened it by simultaneously fitting the three transit signals along with a variability model, similar to the methods used by \citet[][albeit without also simultaneously fitting for systematic errors]{Vanderburg2016} and \citet{pepper}. We use this flattened light curve in our MCMC analysis, as described in Section~\ref{fig:MCMCtransitfit}.

\subsubsection{Light Curves for Other Ursa Major Candidate Members}
In addition to kinematic analysis, another important method in distinguishing moving group membership is the analysis of the main sequence candidate population's gyrochronological rotation sequence, as coeval stars will map out a relation between color and rotation period whose slope is dependent upon the cluster age. We refer the interested reader to \cite{Meibom2011} for a more in-depth discussion of this topic. 
Candidate UMa members were identified in a volume-limited search (50\,pc radius) of \name{} and further assessed for membership using the process described in Section~\ref{subsec:Uma}.

To measure photometric rotation periods of UMa candidates, we extracted \tess{} light curves for each target using cutouts of the full frame images (30-minute cadence data for the first two years of observations, 10-minute cadence for the third and fourth years of observations, and 200-second cadence for the fifth and sixth years of observations) via \texttt{TESSCUT}\footnote{\texttt{TESSCUT}: \url{https://mast.stsci.edu/tesscut/}} \citep{tesscut}. 
The light curves were extracted using 20 apertures, and then we performed a systematic correction process similar to that described in Section~\ref{subsubsec:planetsearch} on each light curve. 
For each source, we identified the aperture that produced the light curve with the highest photometric precision. 
All light curves were normalized and sigma clipped using a $3.5\sigma$ threshold.
We then analyzed these light curves to search for photometric rotation periods, a process we describe further in Section~\ref{subsec:Uma}. 

\subsection{Archival Spectroscopy}
\label{sec:recon spec}

As a nearby, bright, young star, \name{} has been the subject of numerous spectroscopic investigations over the decades, and a treasure trove of archival high-resolution spectroscopic observations of the star exists. 
\citet{Mann2020} gathered archival radial velocity observations from the ELODIE \citep{ELODIE}, SOPHIE \citep{Perruchot:2008}, and Lick Hamilton spectrographs \citep{lick}, and collected their high-resolution spectra from the Tillinghast Reflector Echelle Spectrograph (TRES, \citealt{Furesz08}) and the Network of Robotic Echelle Spectrogaphs (NRES, \citealt{NRES2018}). Since the discovery of the two large planets in the system, new spectroscopic observations from HARPS-N \citep{damasso} and CARMENES \citep{mallorquin} have been added to this extensive list. Tabulated versions of the radial velocity observations used in this investigation can be found in Table 2 of \cite{Mann2020}, Table D.1 of \cite{damasso}, and Tables C.1 and C.2 of \cite{mallorquin}. All of these data show no large RV signals corresponding to stellar mass companions, which allows us to rule out a variety of false positive scenarios for our newly detected planet candidate (see Section~\ref{sec:Vetting}).
\subsection{Ground-based Seeing-Limited Photometry}
\label{sec:lcogt}

Signals from blended eclipsing binaries can sometimes result in false positive transit signals --- a concern that is exacerbated by the large pixel size of the \tess{} instrument ($21\arcsec$ per pixel). To rule out such false positive scenarios for \pname{}, we observed the star during the predicted time of transit for the new planet candidate using the Las Cumbres Observatory Global Telescope (LCOGT; \citealt{Brown2013}) 1-meter telescope at McDonald Observatory on 2022 February 24. The observations were conducted with the $4096\times4906$ pixel Sinistro cameras in Sloan $z$ band. The data were reduced by the LCO BANZAI pipeline \citep{banzai}. The light curves were extracted using AstroImageJ \citep{Collins2017} using an aperture radius of $2.5\arcsec$. 
No sources within this distance were identified with AO imaging or in the \Gaia{} data.
While the photometric precision of the target star's light curve was insufficient to detect the shallow transits observed by TESS, the observations showed no evidence of eclipses or other variability in the nearby stars that could be responsible for the signal associated with \pname{} (down to $G_{\mathrm{Gaia}}=19$\,mag). 
For reference, the closest star to our target is a $G_{\mathrm{Gaia}}=19$\,mag star at an angular separation of $18\arcsec$. 
The 2022-02-24 data firmly rule out NEBs in stars within $2.5\arcmin$ of the target over a $-24.3\sigma$ to $+22.2\sigma$ observing window.
All ground-based follow-up light curves are available on ExoFOP.\footnote{\url{https://exofop.ipac.caltech.edu/tess/target.php?id=130181866}}  
%The nearest star that is capable of producing the transit signal is 26\,arcseconds from our target. 

\subsection{Seeing-Limited Archival Photometry}
\label{sec:Deathstar}
To corroborate the non-detection of variability of any stars close to the position of \name{}, we used archival ground-based photometric data from the Zwicky Transient Facility (ZTF) survey \citep{Masci2019} to check whether the \pname{} 4-day transit signal might be the result of a blended eclipsing binary. 
We used the \texttt{DEATHSTAR} software package \citep{deathstar} to download ZTF image cutouts of the sky surrounding \name{} and extract light curves from all the stars within $2.5\arcmin$ of the target. We folded the light curves of each nearby star (8 in total) on the orbital period of \pname\ and ruled out variability that could plausibly cause the transit signal we see in \tess{} from any stars other than \name{}. We conclude that the signal must be coming from \name{}, consistent with the planetary interpretation.

\subsection{Archival Adaptive Optics Imaging}
\label{sec:AO}
To verify the first two exoplanet detections 
in the \name{} system, \cite{Mann2020} collected archival adaptive optics imaging data spanning nearly a decade (1999-2008, \citealt{Mason2001,Raghavan2012}), none of which showed any evidence for bound companions, or any background sources bright enough to cause the transit signals. Combining the detection limits from these archival adaptive optics observations with stellar evolutionary models from \cite{Baraffe2015} based on an inferred system age of 400\,Myr, \cite{Mann2020} ruled out the presence of background sources down to sub-stellar masses. 
We verified this with our own high angular resolution observations: an adaptive optics (AO) observation from Keck Observatory using the NIRC2 instrument on 2020 September 9 in a narrow Br$\gamma$ filter, and a speckle observation from the Southern Astrophysics Research telescope with the HRCam instrument on 2020 December 29.  
For reference, the Keck AO observation have the following parameters: an estimated pixel scale of 0.009942\,arcseconds, PSF of 0.0496482\,arcseconds, and an estimated contrast of 7.310\,mag at 0.5\,arcseconds.
Similarly, the speckle observations have the following parameters: an estimated pixel scale of 0.01575\,arcseconds, PSF of 0.06364\,arcseconds, and an estimated contrast of 7.6\,mag at 1\,arcsecond.
%MSF would be great to add the integration time for both observations.
%MSF checked in with the AO PI to ensure we are not stepping on any toes. 
Neither of these observations showed any new evidence for nearby companions or background objects capable of producing the transit signatures of \pname{}. 

\subsection{Gaia DR3 Data}
We utilized the third data release of the \textit{Gaia} mission (\Gaia{}, \citealt{GaiaDR3}) to investigate the kinematic properties, spatial distribution, and stellar properties of \name{} and a population of candidate UMa members. 
The candidate UMa members were identified and assessed for membership using the processes described in Section~\ref{subsec:Uma}.
\textit{Gaia} magnitudes, colors, parallaxes, extinction values, and reddening constraints were employed to determine the properties of this population of targets. 

\subsection{Archival Lithium Abundances} \label{sec:archivallithium}
The published lithium abundance measurements for \name{} and the identified UMa catalog sources are drawn from the following catalogs: \cite{Soderblom1993,King2005,Mishenina2008,Ammler2009,Gonzalez2010,Ramirez2012,Luck2017,Aguilera2018,Llorente2021}. 
We found that fewer than 20\% of our catalog members had corresponding lithium abundance measurements in the literature. 

\section{Host Star Properties}\label{sec:properties}
\subsection{Properties of the Host Star}

\name{} is a well-studied star thanks to its proximity to Earth and long-suspected membership of the Ursa Major moving group, and was thoroughly characterized by \citet{Mann2020}. Since that work, an updated parallax from \textit{Gaia} DR3 has been released, but the change does not appreciably impact the stellar parameters, so we simply adopt the stellar parameters from \citet{Mann2020} for the rest of this work. 
The directly observed and inferred properties of the host star \name{} (TIC~130181866, TOI~1726, HIP~38228, \textit{Gaia} DR3 875071278432954240) are shown in Table~\ref{tab:prop}, and the abundance properties are provided in Table~\ref{tab:abundanceprop} (from the Hypatia Catalog, \citealt{Hypatia}).
\begin{deluxetable}{lccc}[htb!]
\label{sec:host_star}
\centering
\tabletypesize{\scriptsize}
\tablewidth{0pt}
\tablecaption{Properties of the host star \name. \label{tab:prop}}
\tablehead{\colhead{Parameter} & \colhead{Value} & \colhead{Source} }
\startdata
TIC ID & 130181866 & TESS Input Catalog\\
TOI ID & 1726 & \cite{Guerrero2021}\\%\url{https://tev.mit.edu/data/}\\
\textit{Gaia} DR3 ID &  875071278432954240 & \Gaia{}\\
%& HD 63433 & \cite{Cannon1993}\\
\hline
\multicolumn{3}{c}{Astrometry}\\
\hline
$\alpha$.  & 07:49:55.048 & \Gaia{} \\
$\delta$. & +27:21:47.28 & \Gaia{} \\
$\mu_\alpha$  & $-10.220 \pm 0.022$ & \Gaia{} \\
$\mu_\delta$ (mas\,yr$^{-1}$) & $-11.235 \pm 0.014$ & \Gaia{}\\
$\pi$ (mas) & $44.6848\pm0.0228$ & \Gaia{}\\
distance (pc) & \dist{} & \Gaia{}\\
\hline
\multicolumn{3}{c}{Photometry}\\
\hline
Spectral Type & G5V & \cite{Gray2003} \\
G$_{Gaia}$ (mag) & \gaiamag{} & \Gaia{}\\
%describes why a Gmag error is not a single value https://gea.esac.esa.int/archive/documentation/GDR2/Gaia_archive/chap_datamodel/sec_dm_main_tables/ssec_dm_gaia_source.html
BP$_{Gaia}$ (mag) & $7.0763 \pm 0.0034$ & \Gaia{}\\
RP$_{Gaia}$ (mag) & $6.2228 \pm 0.0042$ & \Gaia{}\\
B$_T$ (mag) & $7.749 \pm 0.016$ & Tycho-2 \\
V$_T$ (mag) & $6.987 \pm 0.010$ & Tycho-2\\
J (mag) & $5.624 \pm 0.043$ & 2MASS\\
H (mag) & $5.359 \pm 0.026$  & 2MASS\\	
Ks (mag) & $5.258 \pm 0.016$ & 2MASS\\
W1 (mag) & $5.246 \pm 0.178$ & ALLWISE\\
W2 (mag) & $5.129 \pm 0.087 $ & ALLWISE\\
W3 (mag) & $5.297 \pm 0.016$ & ALLWISE\\ 
W4 (mag) & $5.163 \pm 0.031$ & ALLWISE\\ 
\hline
\multicolumn{3}{c}{Kinematics \& Galactic Position}\\
\hline
RV$_{\rm{Bary}}$  (km\, s$^{-1}$) & $-16.07 \pm 0.13$ & \Gaia{}\\
%RV$_{\rm{Bary}}$ (km\, s$^{-1}$) & -15.81$\pm$0.10 & \cite{Mann2020} \\
%U (km\, s$^{-1}$) & 13.66$ \pm $0.09 & \cite{Mann2020}\\
%V (km\, s$^{-1}$) & 2.42$ \pm $0.02 & \cite{Mann2020}\\
%W (km\, s$^{-1}$) & -7.75$ \pm $0.04 & \cite{Mann2020}\\
U (km\, s$^{-1}$) & $13.88 \pm 0.09$ & This paper \\
V (km\, s$^{-1}$) & $2.49 \pm 0.02$ & This paper \\
W (km\, s$^{-1}$) & $-7.87 \pm 0.04$ & This paper \\
X (pc) & $-19.84 \pm 0.03 $ & This paper \\
Y (pc) & $-4.68 \pm 0.01 $ & This paper \\
Z (pc) & $9.14 \pm 0.01 $ & This paper \\
%X (pc) & -19.89$ \pm $0.02 & \cite{Mann2020}\\
%Y (pc) & -4.697$ \pm $0.005 & \cite{Mann2020}\\
%Z (pc) & 9.164$ \pm $0.091 & \cite{Mann2020}\\
\hline
\multicolumn{3}{c}{Physical Properties}\\
\hline
$P_{\rm{rot}}$ (days) & $6.4 \pm 0.6$ & This paper\\
$L_X/L_{\rm{bol}}$ & {$(9.1 \pm 2.7)\times10^{-5}$} & \cite{Mann2020}\\
$\log R'_{\rm{HK}}$ & $-4.39\pm0.05$ & \cite{Mann2020}\\
%\vsini \ (km\,s$^{-1}$) & $ 7.3\pm0.3 $ & \cite{Mann2020}\\
\vsini \ (km\,s$^{-1}$) & $ 7.26 \pm 0.15 $ & \cite{Mallorquin2023}\\
$i_*$ ($^\circ$) & $ >71$ & \cite{Mann2020}\\
%\fbol\,(erg\,cm$^{-2}$\,s$^{-1}$)& \tbd{($4.823\pm0.12)\times10^{-8}$} & \cite{Mann2020}\\ 
Age (Myr) & \age{} & \cite{Jones2015} \\
T$_{\mathrm{eff}}$ (K) & $5688 \pm 28$ & \cite{Hypatia} \\
%M$_\star$ (M$_\odot$) & $0.99 \pm 0.03$ & \cite{Mann2020} \\
%$\mathrm{M}_\star$ (\Msun{}) & $0.9675 \pm 0.085$  \\
$\mathrm{M}_\star$ (\Msun{}) & \mass{} & \cite{Mann2020} \\
R$_\star$ (\Rsun{}) & \rad{} & \cite{Mann2020}\\ 
%L$_\star$ (\Lsun{}) & $0.76 \pm 0.04$ & This paper\\
%MSF computed this using the error propagation from T 
L$_\star$ (\Lsun{}) & $0.75 \pm 0.03$ & \cite{Mann2020} \\
$\rho_\star$ ($\rho_\odot$) & $1.3 \pm 0.2$ & \citet{Mann2020} \\
$\log{(g)}$ ($\log(\mathrm{cm/s^2})$) & $4.52 \pm 0.05 $ & \cite{Hypatia} \\
\enddata
\end{deluxetable}

\begin{deluxetable}{lcc}[tb!]
\centering
\tabletypesize{\scriptsize}
\tablewidth{0pt}
\tablecaption{Abundance Properties of \name. All values were obtained using the Hypatia Catalog. We report only sources with errors less than 0.3\,dex \citep{Hypatia}.} 
\label{tab:abundanceprop}
\tablehead{\colhead{Parameter} & \colhead{Value}}
\startdata
$[$Fe/H$]$ (dex) & \feh{}  \\
A(Li) (dex) & $2.5 \pm 0.2$  \\
\hline
$[$C/H$]$ (dex) & 	$-0.09	\pm 0.074$  \\
$[$N/H$]$ (dex) & 	$-0.22 \pm	0.11$ \\
$[$O/H$]$ (dex) & $0.00	\pm 0.014$  \\
$[$Na/H$]$ (dex) & $-0.08 \pm 0.24$  \\
$[$Mg/H$]$ (dex) & 	$0.02 \pm 0.19$  \\
$[$Al/H$]$ (dex) & 	$-0.05 \pm 0.19$ \\
$[$Si/H$]$ (dex) & $-0.03\pm 0.08$  \\
$[$SiII/H$]$ (dex) & $0.05\pm0.05$  \\
$[$S/H$]$ (dex) &	$0.18\pm0.09$  \\
$[$Ca/H$]$ (dex) &	$0.09\pm0.086$  \\
$[$CaII/H$]$ (dex) & $0.56\pm0.03$  \\
$[$Sc/H$]$ (dex) &	$0.15\pm0.067$  \\
$[$ScII/H$]$ (dex) & $0.08\pm0.09$  \\
$[$Ti/H$]$ (dex) &	$0.03\pm0.17$  \\
$[$TiII/H$]$ (dex) & $0.13\pm0.06$  \\
$[$V/H$]$ (dex) & $0.00\pm0.094$  \\
$[$VII/H$]$ (dex) &	$0.06\pm0.11$  \\
$[$Cr/H$]$ (dex) &	$0.04\pm0.094$  \\
$[$CrII/H$]$ (dex) & $-0.24\pm0.07$  \\
$[$Mn/H$]$ (dex) &	$-0.15\pm0.16$  \\
$[$Co/H$]$ (dex) &	$0.07\pm0.04$  \\
$[$Ni/H$]$ (dex) &	$-0.04\pm0.10$  \\
$[$Zn/H$]$ (dex) &	$-0.34\pm0.16$ \\
$[$Zr/H$]$ (dex) &	$0.28\pm0.10$  \\
$[$PrII/H$]$ (dex) & $0.31\pm0.10$  \\
$[$Eu/H$]$ (dex) &	$-0.04\pm0.11$  \\
$[$EuII/H$]$ (dex) & $-0.04\pm0.28$  \\
\enddata
\end{deluxetable}
Most notably, \name{} is bright ($V\simeq6.9$\,mag), young (\age{}\,Myr), and positioned at a close distance from the Earth (\dist{}\,pc). 
Additionally, \name{} is a solar analog (spectral type G5V), exhibiting solar-like qualities, including a mass of \mass{}\,\Msun{} and $[\textrm{Fe}/\textrm{H}] = $~\feh{}\,dex \citep{Hypatia}. 
Due to the youth of the host star, it differs substantially from the Sun in terms of its strong lithium abundance (A(Li)~$=2.5\pm0.2$\,dex), short rotation period ($P_{\mathrm{rot}}=6.4\pm0.6$\,d), and high X-ray luminosity ($L_X/L_{\rm{bol}}=0.75\pm0.03$). 

The Galactic space velocity values were calculated using \texttt{PyAstronomy} tools that draw from the methodology described in \cite{Johnson1987}. 
These calculations relied on \Gaia{} kinematic and positional measurements.
{The spatial center of the UMa moving group is $(X,Y,Z)=(-7.5,+9.9,+21.9)$\,pc \citep{Gagne2018}. This can be compared to the Cartesian Galactic coordinates of \name{}, measured as $(x', y', z') = (-19.89, -4.70, +9.16)$\,pc.
We determined that \name{} is approximately 23\,pc the central point of the UMa moving group. This is similar to the Earth-\name{} distance, but in the opposite direction, making HD 63433 roughly equidistant between the Solar System and UMa core.}
This estimate is in good agreement with the value determined in \cite{Mann2020}.  

{Our analysis incorporates stellar properties determined by \citet{Mann2020}. They fit the spectral-energy distribution using a grid of empirical templates (with atmosphere models to fill in wavelength gaps). The method provides two estimates of $R_{\star}$, one from the Stefan-Boltzmann relation (using the \Gaia{} parallax and bolometric flux from the SED) and another from the infrared-flux method \citep{1977MNRAS.180..177B}, i.e., from the scale factor between the absolutely calibrated spectrum and the model spectrum. Both radii agree, and both methods reproduce radii from long-baseline interferometry \citep[e.g.,][]{2012ApJ...753..171V}. The resulting $T_{\rm eff}$ and $R_*$ are within $1\sigma$ of those reported from \Gaia{} \citep{2022yCat.1355....0G}.
The stellar radius uncertainty from \citet{Mann2020} was only propagated along with the mass uncertainty for the starting parameter for stellar density, but $R_{\rm p}/R_{\star}$ was a separate parameter that had its own value and uncertainty also determined from the transit models in \citet{Mann2020}.}

\section{Analysis}\label{sec:analysis}
\subsection{Transit Analysis \& Results}
\label{sec:methods}

We used the \textit{Easy Differential-Evolution Markov Chain Monte Carlo algorithm},  \texttt{edmcmc} \citep{vanderburgedmcmc},\footnote{\url{https://github.com/avanderburg/edmcmc}} to obtain transit fit parameters and uncertainties for the three planets in the \name{} system \citep{terbraak2006}. 
The method involves generating many possible solutions, or model parameters, that describe the light curve transit signal. 
These parameters are sampled from a probability distribution using the Markov Chain Monte Carlo (MCMC) technique.
%The differential evolution algorithm moves the MCMC chain from one parameter set to another, generating new proposals based on the difference between randomly selected chains. 
This technique helps to ensure that the chain explores the full parameter space and prevents the algorithm from getting stuck at local minima.

We then employed the BAsic Transit Model cAlculatioN \texttt{Python} package  (\texttt{BATMAN}; \citealt{batman}), producing three transit models that the \texttt{edmcmc} then fit simultaneously. 
For reference, \texttt{BATMAN} requires inputs for the following variables: time of inferior conjunction ($T_0$), planet orbital period ($P$), planet-to-star radius ratio ($R_p/R_*$), inclination ($i$), argument of periastron ($\omega$), eccentricity ($e$), limb-darkening coefficients (described below), and scaled semi-major axis ($a/{R_\star{}}$).

To reduce the number of input parameters into the \texttt{edmcmc}, each planet's semi-major axis ($a/{R_{\star}}$) was calculated by leveraging Kepler's third law (under the assumption that $M_{\mathrm{p}}/M_{\star}\ll1$),
\begin{equation}
\frac{a}{R_{\star}}=\sqrt[3]{\frac{G P^2 \rho_*}{3\pi}}, 
\end{equation}
where $G$ is the gravitational constant, $P$ is the orbital period, and $\rho_*$ is the mean stellar density. 
This allowed us to replace the orbital separations of individual planets with a single free parameter (the mean stellar density).
We also re-parameterized the eccentricity and argument of periastron for each planet as $\sqrt{e} \sin(\omega)$ and $\sqrt{e} \cos(\omega)$, which has been shown to improve MCMC convergence. 
We also fit for linear and quadratic limb-darkening coefficients ($u_1, u_2$), which we fit for using the ($q_1, q_2$) parameterization described by \citet{Kipping2013}. In order to interface with \texttt{BATMAN}, which calculates models in terms of  $u_1$ and  $u_2$, we solved for these coefficients using the following transformations: 
\begin{equation}
u_{1}=2 \sqrt{q_{1}} q_{2}
\end{equation}
\begin{equation}
u_{2}=\sqrt{q_{1}}\left(1-2 q_{2}\right),
\end{equation}
as determined by \cite{Kipping2013}.

For our \texttt{edmcmc} analysis, we imposed Gaussian priors on the mean stellar density and limb-darkening coefficients. The priors imposed on the limb-darkening coefficients are the same as those used by \citet{Mann2020}. The width of the stellar density prior was found via error propagation using the stellar mass ($M_\star$) and radius ($R_\star$) uncertainties calculated by \cite{Mann2020} after our derivation of the parameter. 
Other parameters are sampled uniformly, some using physically-motivated constraints, such as the optimal impact parameter ($b$) for each planet being constrained to $|b|<1+R_P/R_\star$, $R_p/R_\star>0$, and $\sqrt{e}\cos\omega$ and $\sqrt{e}\sin\omega$ were restricted to (-1,1). Lastly, we restricted the eccentricity of the three planet orbits to be $e < 0.9$, because high values of eccentricity can sometimes lead to failures of \texttt{BATMAN}'s Kepler equation solver. This did not appreciably impact the posterior probability distributions for eccentricity, as the transit data were sufficient to constrain eccentricity to be less than this value. 
The priors implemented for all three planets are listed in Table~\ref{tab:priors}.
\begin{table}[tbh!]
\centering
\begin{tabular}{cc}
Parameter & Prior \\ \hline
$T_0$ (TJD) & $\mathcal{U}[-\infty,\infty]$ \\
$P$ (days) & $\mathcal{U}[0,\infty]$ \\
$R_P/R_{\star}$ & $\mathcal{U}[0, 1]$ \\
$b$ & $\mathcal{U}[|b|<1+R_P/R_*]$  \\
$\rho_{\star}$ ($\rho_{\odot}$) & $\mathcal{N}[1.312, 0.152]$ \\
$q_{1,1}$ & $\mathcal{N}[0.30,\, 0.06]$ \\
$q_{2,1}$ & $\mathcal{N}[0.37, 0.05]$ \\
$\sqrt{e}\sin\omega$            & $\mathcal{U}[-1,1]$ \\
$\sqrt{e}\cos\omega$            & $\mathcal{U}[-1,1]$ \\ \hline
\end{tabular}
\caption{$\mathcal{U}[X,Y]$ indicates a uniform prior with limits $X$ and $Y$, and $\mathcal{N}[X,Y]$ indicates a Gaussian prior with mean $X$, and standard deviation $Y$.\label{tab:priors}}
\end{table}

Key parameters in the \texttt{edmcmc} code include the number of walkers (the number of individual chains), the number of links (the number of steps that an individual chain will take), and the number of burn-in steps.
For context, the number of burn-in steps represents the number of cycles in the initial portion of the MCMC chain that is to be discarded, as the algorithm has yet to reach equilibrium and could bias the evolution of the posterior distribution.
Each run of the \texttt{edmcmc} code returns an object, which contains the positions of each walker at each link.

We initialized the MCMC chains at parameters near the values reported for \name{}\,b and \name{}\,c by \cite{Mann2020}.
In contrast, we initialized the parameters for \pname{} based on a by-eye inspection of the BLS power spectrum and phase-folded light curve of the period corresponding to the strongest peak in the power spectrum.
Given that transiting exoplanets exhibit a near-edge on inclination, we set the initial inclination input for \pname{} to $90^\circ$.

For our initial run, the eccentricity ($e$) and argument of periastron ($\omega$) were fixed to $0$.
By running our model with $e=\omega=0$, we simplify our procedure by reducing the number of free parameters to be explored.
This initial run with $e=\omega=0$ employed $10^2$ walkers, $5 \times 10^4$ links, and $10^3$ burn-in steps. We utilize the well-converged results from this run as the initial positions for our subsequent runs, where $e$ and $\omega$ were then incorporated as free variables. For these runs which explore a larger parameter space, we employ $10^2$ walkers, $10^6$ links, and $10^4$ burn-in steps.

To determine if the \texttt{edmcmc} code has converged to the true posterior distribution of our system, we employed the \cite{GelmanRubin} convergence statistic.
For reference, a Gelman-Rubin value near unity (largest value of 1.2) for each parameter is an indicator that convergence has been achieved.
To improve upon the convergence of our first run, we incorporated the results from this run as the starting position for a second run with $10^6$ links. We repeated this process without applying burn-in steps for the successive runs, as the starting positions are taken from the preceding runs. 
We were ultimately satisfied with the convergence of the posterior distribution parameters after repeating this step seven times, resulting in a combined chain of seven million links. We then used our well-converged \texttt{edmcmc} outputs to construct transit models for each of the three detected \name{} planets, once again employing the \texttt{BATMAN} package. 

\begin{figure*}[tbh!]
    \centering
    \includegraphics[width=0.98\textwidth]{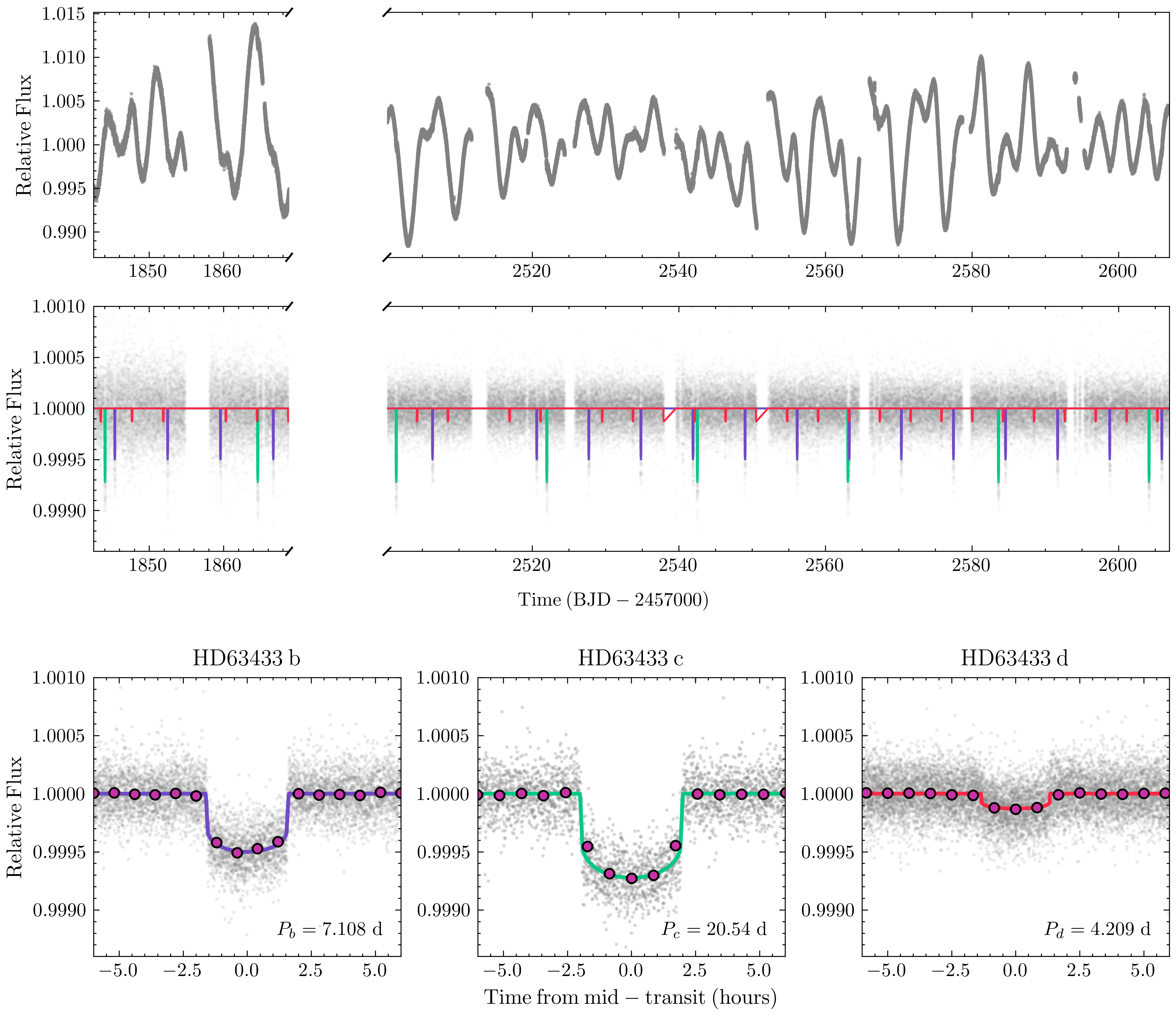}
    \caption{The top panel shows the unflattened TESS light curve of \name{}. The middle panel shows the flattened light curve with our three best-fit transit models for planets b, c, and d, as determined by our \texttt{edmcmc} outputs. The colors of these models correspond to each of the three planets as shown in the bottom panels, which depict the phase-folded light curves for each \name{} planet with binned data (red points) plotted over our best-fit models.}
    \label{fig:MCMCtransitfit}
\end{figure*}

The resulting best-fit parameters and uncertainties from our \texttt{edmcmc} investigation are listed in Table~\ref{tab:transfit}. 
Figure~\ref{fig:MCMCtransitfit} depicts the unflattened \name{} \tess{} light curve (top panel), the flattened light curve with the three \texttt{BATMAN} transit models overlaid (middle panel), and the phase-folded light curves and transit models for the three planets (bottom panel). These results are discussed further in Section~\ref{sec:results}.

\subsection{Association of HD 63433 with Ursa Major Moving Group}\label{subsec:Uma}

\name{} was previously determined to be a likely member of the UMa moving group via an analysis of kinematics, lithium abundance, and stellar rotation \citep{Mann2020}. Now, with the release of \textit{Gaia} DR3, it is possible to better characterize the Ursa Major moving group, identify new likely members, and assess whether \name{}'s membership remains solid. 
Therefore, we perform a moving group membership analysis, drawing from a volume-limited candidate population of UMa members. In this section, we describe the candidate member identification process and the methods used to cull this candidate list down to likely members. We note that while other groups have also investigated the membership of Ursa Major using Gaia data \citep[e.g.][]{gagne2020}, we choose to perform our own identification of possible sources for consistency. 

\subsubsection{Identifying UMa Candidates}
To identify a volume-limited population of co-moving stellar companions for the \name{} target, our team employed the open-source \texttt{Python} tool \texttt{Comove}\footnote{\url{https://github.com/adamkraus/Comove}} \citep{toffelmire2021}.
The software identifies co-moving sources with the procedure outlined below.
\begin{itemize}
    \item The software queries the \textit{Gaia} DR3 catalog drawing targets from a volume radius equivalent to the defined search sphere ($r=50$\,pc in our case), and with tangential velocities (drawing from the target's proper motion) within a specified input velocity difference range ($\pm 5$\,km/s in our case).
    \item The software queries the 2MASS, GALEX, ROSAT, and WISE catalogs, recording measurements for the identified kinematic companions.
    \item The software outputs a table listing parameters corresponding to the candidate population. Among the parameters reported are the following: the observed radial velocity from \textit{Gaia} DR3 (when available), the radial velocity value predicted if the source were co-moving with the target with the given tangential velocities, the angular/linear separation between a given source and the target, and \textit{Gaia} DR3 Renormalised Unit Weight Error (RUWE), which can be used to flag for potential binary companions, \citep[e.g.,][]{rizzuto, evans, belokurov}.  
\end{itemize}

\texttt{Comove} derived a table of 1180 potential co-moving candidates.
We then culled this table to only include sources with reported \textit{Gaia} DR3 radial velocity measurements that were within $\pm5$\,km/s of \texttt{Comove}-predicted value. 
This step resulted in the greatest removal of targets, bringing our table of potential co-moving candidates down to 290 sources. 
It is important to note that 407 of the original candidates did not possess reported \textit{Gaia} DR3 radial velocity measurements, due to the faintness and brightness limitations of the current radial velocity survey.
This limited UMa candidates to a population of main sequence stars spanning masses of $0.2-1.5$\,\Msun{}, as determined by PAdova and TRieste Stellar Evolution Code (PARSEC, version 1.2S) \citep{parsec2012}).
The sources removed at this step warrant further study if/when radial velocity measurements are made available.
Next, we removed sources that have not yet been observed by TESS, further reducing our table of potential co-moving candidates to 130 sources. 

\subsubsection{Measuring Photometric Rotation Periods for HD 63433 and Our UMa Candidates}
We investigated the photometric rotation periods for all 130 UMa candidates.
To determine the best period corresponding to a given target, each source was analyzed using three period-search techniques.
This included a Lomb Scargle (LS) periodogram analysis \citep{Lomb1976, Scargle1982, Zechmeister2009}; the Phase Dispersion Minimization (PDM) algorithm \citep{pdm}; and an autocorrelation function (ACF, \citealt{mcquillan}).

These three period-search techniques were implemented on light curves from each individual sector, while the LS and PDM techniques were also implemented on a full light curve, comprised of all available sectors stitched together. 
The ACF technique was applied only to individual sectors because the large gaps between \tess{} observations adversely affected the results. 

We searched for periodic signatures within a period range of 0.04\,d (twice the integration time) to 30\,d.
If the observational baseline ($\tau$) for a given source was less than 90\,d, we reduced the maximum period search limit to $\tau/3$.
We performed a by-eye analysis of all resulting power spectra, autocorrelation plots, and phase-folded light curves corresponding to the strongest three periods of each of the periodograms. 
Two individuals from our collaboration team inspected the periodogram results for each candidate to search for signs of a periodic signature and determine the corresponding rotation period.
When discernible, the photometric rotation periods for a given source were cataloged along with an accompanying error estimate, which we estimated by taking the larger of either the error calculated by the PDM periodogram \citep{Schwarzenberg1997} or 10\% of the rotation period (to account for systematic uncertainties in the rotation period estimates).  

In some cases, we observed that \tess{} instrumental systematics on the timescale of the spacecraft's orbit obscured very short-period rotation signals. To better recover the rotation periods in these cases, we then searched the light curves with all three methods (LS, PDM, and ACF) after high-pass-filtering the light curve by fitting a B-spline with break-points spaced every few days to the light curve and subtracting out the best fit and adding 1 (to return to the original normalization scale). We also introduced discontinuities in the B-spline at the beginning of each new \tess{} orbit to better model the rapid systematic changes in flux due to scattered light from the spacecraft's close passage near Earth. Performing this additional analysis on the high-pass-filtered light curve led to the identification of several additional rotators, mostly with very short rotation periods (less than 2 days). 

We measure photometric rotation periods for 56 of the 130 stars, which we list in Table~\ref{tab:UMaG_rot}.
We refer to these 56 sources as UMa members from this point forward.
Aside from our planet host, which is presented in bold typeface in the top row, sources are listed in ascending order by reddened-corrected color. 
In all cases, the reddening correction was less than 1\% of the total BP-RP.
The table also provides the TIC ID, spatial coordinates, apparent and absolute \textit{Gaia} magnitudes, apparent and absolute colors, distance from Earth, and RUWE value.
Our cataloged UMa members are all bright, nearby targets, ranging between 12-57\,pc from the Earth and 14.6-4.5\,mag in apparent \textit{Gaia} magnitude. 

\subsubsection{Rotation Sequence Consistency Among UMa Members}
\label{subsubsec:rot}
The correlation between color and stellar rotation period (gyrochronological sequence) in moving groups has been well documented \citep[e.g.,][]{2007ApJ...669.1167B, Hartman2010,vanSaders2016, Douglas2016}.
This relation builds upon the foundational work of \cite{Skumanich1972}, who discovered a correlation between the average rotational velocity (\vsini{}) and the age of a given stellar cluster.

\begin{figure*}[t!]
    \centering
    \includegraphics[width=\textwidth]{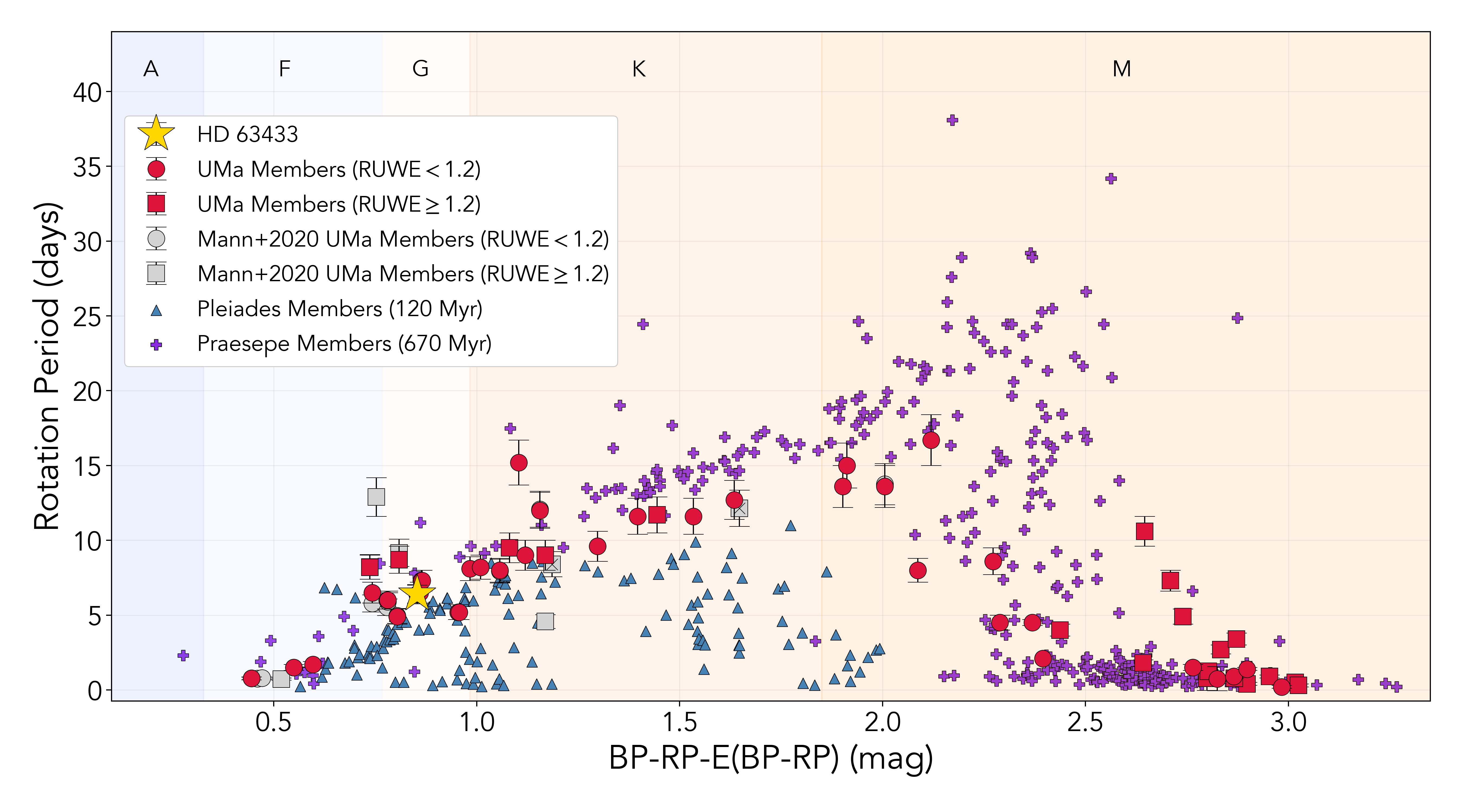}
    \caption{The color-rotation sequence for catalog members of the UMa moving group, as compared to the 120\,Myr Pleiades \citep{Rebull2016} cluster and the 670\,Myr Praesepe cluster \citep{Douglas2017}. 
    Circles indicate UMa sources with RUWE~$<1.2$, while squares indicate those with RUWE~$\geq1.2$. 
    Our target \name{} is shown as a gold star. 
    Light blue circles/squares denote UMa sources featured in \cite{Mann2020} --- those shown with an `x' symbol overplotted lack \Gaia{} reddening values required for absolute calibration.
    We identify the spectral types investigated, adding shaded panels that correspond to the hexadecimal color codes for a given spectral type  (Luminosity class V, subclass 5), as reported by \cite{Harre2021}.}
    \label{fig:umajor1}
\end{figure*}

We illustrate the color-rotation sequence of our UMa catalog members in Figure~\ref{fig:umajor1}.
We annotate the top of the figure to indicate the range of spectral types, adding shaded panels that correspond to the hexadecimal color codes for a given spectral type  (corresponding to Luminosity class V and subclass 5), as reported by \cite{Harre2021}.
The gyrochronological relation only holds for main sequence stars, however, as we will show in Section~\ref{subsubsec:cmd}, our UMa catalog members are all positioned along the main sequence.
Therefore, we include the full UMa catalog in Figure~\ref{fig:umajor1}. 
UMa catalog sources that are likely isolated (RUWE~$<1.2$) are denoted as crimson circles, while potential binaries (RUWE~$\geq1.2$) are depicted as crimson squares (both contain error bars). 
UMa sources reported in \cite{Mann2020} but not included in our catalog are shown as light blue squares/circles (the same RUWE criterion applies here). 
We also plot members of the 670\,Myr Praesepe cluster \citep{Douglas2017} and the 120\,Myr Pleiades cluster \citep{Rebull2016}, incorporating the \textit{Gaia} DR3 reddening-corrected color to optimally align the three systems.
In cases where the reddening corrections were not available (such as the small subset of \cite{Mann2020} UMa catalog sources not included in our catalog), we annotate symbols with an ``x" symbol. 

The UMa sources form a well-defined color-rotation sequence, with bluer, more massive UMa members exhibiting shorter periods. For redder UMa members, the rotation period increases for lower mass stars until the tight sequence ends for M-dwarf members, and the rotation periods are more randomly distributed. Importantly, our planet-hosting star, \name{}, falls well within the UMa color-rotation sequence, corroborating its association with the moving group. It is shown in the plot as a gold star. 

Our updated color-rotation sequence has several additional interesting features. We measure a rotation period of 0.77 days for the bluest UMa catalog source (F1V spectral type), which is well aligned with the rotation periods of F1V stars in other young clusters, including the two shown in Figure~\ref{fig:umajor1}.
The slope of the color-rotation relation and the corresponding point of divergence (the mass below which there is no tight rotation sequence) are age-dependent, with older clusters exhibiting steeper slopes and divergences from the relation at lower masses.
We observe that for our UMa catalog members, the divergence from the color-rotation relation occurs for sources later than spectral type M1V, as would be expected for a cluster with an age estimate of \age{}~Myr \citep{Jones2015}.
In contrast, this divergence occurs at spectral type K4V for the younger Pleiades cluster and at later M-type for the older Praesepe cluster. UMa's color-rotation relation for FGK stars also shows a slope intermediate to that of the Pleiades and Praesepe systems, corroborating the fact that UMa's age is between that of those two clusters. 

\subsubsection{Color-Magnitude Diagram Consistency Among UMa Members}
\label{subsubsec:cmd}
\begin{figure}[tbh!]
    \centering
    \includegraphics[width=0.48\textwidth]{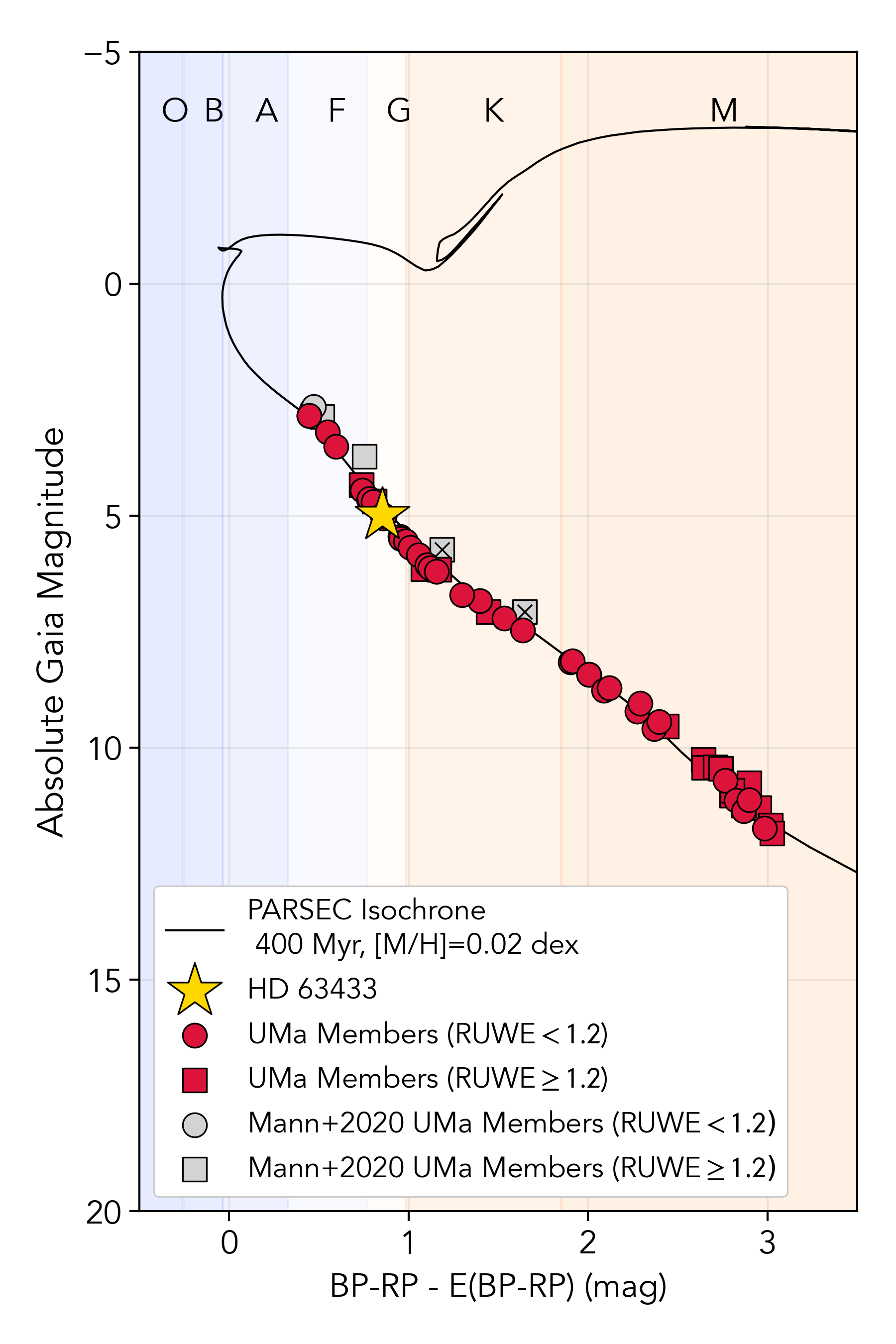}
    \caption{Color-magnitude diagram of \name{} (gold star) and our catalog of UMa co-moving members (crimson circles/squares). 
    Same legend definitions as Figure~\ref{fig:umajor1}.
    In black, we illustrate a PAdova and TRieste Stellar Evolution Code (PARSEC) isochrone (version 1.2S) for a 400\,Myr system with [M/H]=0.02\,dex \citep{parsec2012}. 
    We identify the spectral types investigated, adding shaded panels that correspond to the hexadecimal color codes for a given spectral type  (Luminosity class V, subclass 5), as reported by \cite{Harre2021}.}
    \label{fig:umajor2}
\end{figure}

We also assessed whether HD 63433's color-magnitude diagram position is consistent with that of the other UMa catalog members. We find that our UMa catalog members are well-matched to a 400\,Myr, [M/H]=0.02\,dex stellar isochrone, as shown by the result drawn from the PAdova and TRieste Stellar Evolution Code (PARSEC, version 1.2S) in Figure~\ref{fig:umajor2} \citep{parsec2012}.\footnote{\url{http://stev.oapd.inaf.it/cgi-bin/cmd}} We find that \name{} (shown as a gold star in Figure~\ref{fig:umajor2}) falls precisely in the sequence of other UMa members in color-magnitude space, in close agreement with the PARSEC isochrone prediction, as expected for single star member of the group.  

The fact that the UMa candidate members closely match a 400 Myr isochrone is encouraging for the fidelity of the sample, and suggestive that the stars are coeval, but because of the spectral types of our candidate members, this is not a particularly stringent test. As described in Section~\ref{sec:intro}, the CMD location of GKM main sequence stars is consistent with isochrones spanning several billion years, therefore, only sources with strong departures from the isochrone can be flagged as potential field candidates. Such departures may arise among main sequence stars with appreciably different metallicity but also may arise from binarity and/or nonstandard stellar evolution. 

We do see some such deviations from the color-magnitude sequence for a handful of candidate UMa members in Figure~\ref{fig:umajor2}. In particular, a handful of sources included in the \cite{Mann2020} analysis, but excluded from our list of candidate members (shown as blue circles/squares), show significant departures from the color-magnitude diagram (CMD). We attribute these deviations from the color-magnitude sequence to the effects of binarity, where the combined light from blended binary stars causes the unresolved source to sit above the color-magnitude sequence in absolute magnitude. 
Supporting the binary hypothesis is the fact that 4/5 of these stars were excluded from our sample because of their high RUWE metrics (RUWE~$>2$).

We note that other possible causes for deviations from a tight color-magnitude sequence include a range of spot filling factors (that are not included in models) or incorrect or missing corrections for extinction and reddening. We suspect the former explanation is more likely because reddening corrections are negligible for UMa members due to their proximity. The average E(BP-RP) reddening correction for our UMa catalog sources is $0.002\pm0.001$\,mag, much less than the uncertainties on these quantities. We label sources missing these corrections with an ``x'' symbol.  

We annotate the top of Figure~\ref{fig:umajor2} to indicate the range of spectral types included on the color-magnitude diagram, adding shaded panels that correspond to the hexadecimal color codes for a given spectral type  (Luminosity class V, subclass 5), as reported by \cite{Harre2021}.
Our UMa catalog members range in spectral type between F1V-M5V, which corresponds to a mass range of \maxmass{}-\minmass{}\,\Msun{} (as determined by a fit to the PARSEC isochrone).

For the sake of completeness and to include more evolved/luminous stellar members in this catalog, we searched for kinematic matches among subclass IV targets and main sequence stars of earlier spectral types that were included in prior UMa catalogs (before the kinematic precision made available by the \textit{Gaia} mission), drawing from resources such as \citet{Miczaika1948} and \citet{Soderblom1993}. 
Such sources could help to establish a more well-defined main sequence turnoff point, which would permit the calculation of a main sequence turnoff mass and corresponding age for the UMa system.
However, we were unable to identify any kinematic matches. 
Part of the difficulty in our search for more luminous kinematic members is because \textit{Gaia} DR3 radial velocity measurements were often not available for these sources. 
A targeted radial velocity follow-up study would offer the potential to identify more luminous kinematic UMa members.

\subsubsection{Lithium Sequence Consistency Among UMa Members}
\label{lithium}
\begin{figure}[tbh!]
    \centering
    \includegraphics[width=0.48\textwidth]{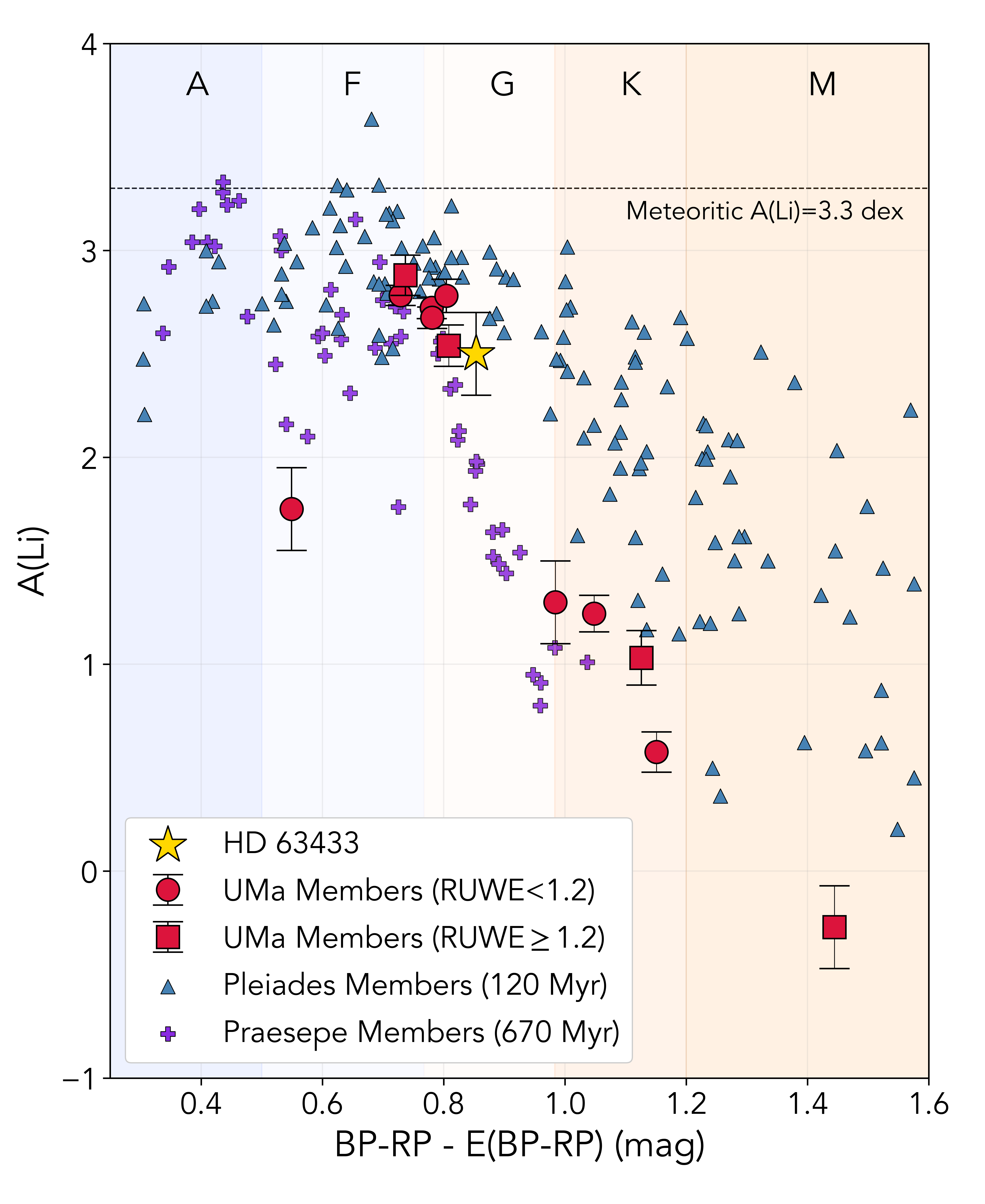}
    \caption{Color-A(Li) sequence of our cataloged UMa members (crimson circles/squares) with \name{} depicted as a gold star. 
    Same legend definitions as Figure~\ref{fig:umajor1}.
    Twelve sources in our catalog had published A(Li) abundances. 
    The solar system meteoritic abundance level of 3.3\,dex is depicted as a horizontal dashed line \citep{Asplund2009}.
    To indicate the range in stellar types, we illustrate the hexadecimal color codes reported by \cite{Harre2021} (Luminosity class V, subclass 5). 
    The color-A(Li) relation for the 120\,Myr Pleiades cluster \citep{Bouvier2018} and the 670\,Myr Praesepe cluster \citep{Cummings2017} are shown as a comparison.
    A/F-type stars in all three systems show evidence of the well-known lithium dip feature (\citealt{Boesgaard1986}). 
    } %
    \label{fig:colorali}
\end{figure}

The abundance of lithium in the stellar photosphere is another important age diagnostic for young stars (\citealt{Jeffries:2014, Barrado:2016}).
Therefore, we searched the literature for measurements of the lithium abundance for our UMa members. We computed a weighted average of the lithium values found in the literature (from the sources described in Section~\ref{sec:archivallithium}). In Figure~\ref{fig:colorali}, these results are illustrated as a function of reddening-corrected color.
To provide a consistent contrast, we illustrate the color-A(Li) relations for members of the Pleiades and Praesepe clusters \citep{Cummings2017, Bouvier2018}.
For the Pleiades and Praesepe members depicted, effective temperatures were transformed to reddening corrected colors using the color conversion function described in \cite{Mucciarelli2021}.

We found that roughly 20\% of our UMa catalog sources have published lithium abundance measurements.
These sources form a clear color-lithium sequence, and the large A(Li) measurements are reflective of a young stellar association. Once again, the lithium abundance for \name{} (shown as a gold star) falls well within the sequence of the other UMa group candidate members, strengthening the case for its association with the group. 
Similar to the color-rotation sequence, UMa's color-lithium relation is shown to map out a region intermediate to that of the Pleiades and Praesepe systems, corroborating the fact that UMa's age is between that of those two clusters. 

In Figure~\ref{fig:colorali},  we include a dashed horizontal line indicating the solar system meteoritic lithium abundance value of 3.3\,dex \citep{Asplund2009}. This is the value corresponding to the initial lithium supply of a given star before the onset of depletion. 
We also note evidence for a lithium dip feature, known to impact A and F-type stars (\citealt{Boesgaard1986}). The departure from a monotonic relation between lithium abundance and color is seen for stars earlier than F5V spectral type, which corresponds to stars with effective temperatures $\gtrsim6600$\,K. 
Our bluest UMa catalog source (F4V spectral type) is shown to coincide with the center of the lithium dip feature, strongly departing from the nearly monotonic color-A(Li) relation traced out by the other UMa catalog sources.

Our updated view of this larger population of UMa members does not change the original conclusion that HD 63433's properties are highly consistent with the group's constituent stars. We, therefore, consider \name{} to be a member of UMa, and apply the group's age measurement of \age{} to \name{} and its planetary companions \citep{Jones2015}.  

\section{Ruling out False Positive Scenarios}% for \pname{}}
\label{sec:Vetting}
\subsection{Possible False Positive Scenarios}
Sometimes, scenarios other than transiting exoplanets can cause apparent dips in the brightness of stars, so transiting exoplanet detections, like that of \pname{}, require scrutiny and analysis before they can be considered solid. One way to increase confidence in exoplanet detections is called ``confirmation," which involves detecting the signal of the planet with an independent detection method. Confirmation often requires expensive follow-up observations, and in many cases is unfeasible given the sensitivity of other detection techniques. Another way to increase confidence in exoplanet detections is ``validation,'' which involves quantifying, testing, and rejecting all other possible false positive scenarios (such as eclipsing binary, blended binary, stellar spots, or instrumental artifacts; see  \citealt{Ciardi2015}), leaving a real exoplanet as the most probable explanation for the candidate transit signal. Because confirmation of \pname{} is likely beyond the capabilities of current precise radial velocity observations (because of HD 63433's high amplitude stellar variability), we use the validation approach to show that \name{} is overwhelmingly likely to be a real exoplanet. 

In particular, we address the following classes of false positive signals and show that \pname{} cannot be caused by each scenario: 

\begin{itemize} 
\item \textbf{Signal arises from instrumental effects and/or stellar variability}: One possibility is that the detected signal is an instrumental artifact or a false alarm resulting from high-amplitude stellar variability. Typically, in such cases, we would anticipate variations in the signal's shape, amplitude, and other attributes across different sections of the light curve. This is because instrumental glitches or stellar variability features are not expected to repeat in perfectly periodic intervals, corresponding to a 4.2\,day orbital period. We test this scenario for \pname{} by assessing the consistency of the transit signal between different transits and different sectors. We found no evidence for variable transit depths. The signal's period is distinct from the stellar rotation period and its harmonics, and there are no known \tess{} instrumental systematics that repeats every 4.2 days. We also confirmed visually that no anomalies in the background flux took place at the times of the transits of \pname\ in any of the observed sectors.  We conclude that the signal is astrophysical and caused by an orbiting object.  

\item \textbf{ HD 63433 itself is an eclipsing binary}: Some stellar eclipses (especially grazing eclipses of small stars) are difficult to distinguish from planetary transits without additional observations, which can be a source of false positives among a list of planetary candidates. It is straightforward to rule this scenario out by collecting radial velocity observations of the host star, as stellar-mass companions will cause Doppler shifts with amplitudes much greater than $1\,\mathrm{km\,s^{-1}}$. Dozens of radial velocity observations collected over the years, and described in Section~\ref{sec:recon spec}, show no velocity shifts greater than about 100\,$\mathrm{m\,s^{-1}}$. Given that the predicted radial velocity semi-amplitude ($K$) values corresponding to an equal mass binary system and a G-type + M-type binary system are much larger than these measurements ($8\times10^{4}\,\mathrm{m\,s^{-1}}$ and $5\times10^4\,\mathrm{m\,s^{-1}}$, respectively), \name{} itself cannot be an eclipsing binary. 

\item \textbf{Contamination by eclipsing binaries}:  \tess{} has a focus-limited PSF with a large pixel scale of approximately $21\arcsec$, so the photometric masks used to extract light curves are large and include flux from several other nearby stars. Thus, these nearby stars that are unresolved by \tess{} but can be resolved by ground-based telescopes can sometimes introduce contaminating signals that mimic planetary transits. 
For context, the number of known \textit{Gaia} sources (including \name{}) within the TESS photometric apertures are as follows: 10 sources in Sector 20, 9 sources in Sectors 44 and 45, 8 sources in Sector 46, and 12 sources in Sector 47.
Using the seeing-limited ground-based observations described in Sections~\ref{sec:lcogt} and \ref{sec:Deathstar}, we placed an upper limit of 6.6\arcsec{} on the radius of confusion regarding this possibility. 
Our observations from the LCO telescope at McDonald Observatory and the archival ZTF light curves we created with \texttt{DEATHSTAR} place a tight upper limit on the radius of confusion among eclipsing binaries with \pname's orbital period. 

In addition to nearby eclipsing binaries far enough from \name{} to be resolved by seeing-limited ground-based telescopes, sometimes false positives are caused by eclipsing binaries close enough to be blended with the target in ground-based images. The vast majority of these targets can be ruled out by adaptive optics observations, which can increase the resolution of ground-based telescopes by orders of magnitude and reveal faint stars that could be the source of the transits.

We quantified how bright any given star must be to cause the transits we observe for \pname{}. Following \citet{Vanderburg2019}, we fit the light curve again with MCMC to measure the duration of transit ingress and egress for this new signal, while imposing no physical constraints {(e.g., not requiring Kepler's third law be obeyed).} We converted this ingress/egress duration into an effective magnitude limit and found that with 99.7\% confidence, for a blended star to mimic HD 63433's transit signal, it must be within 5.57 magnitudes of the target star's brightness. The existing adaptive optics imaging rules out stars this bright at distances greater than about 150\,mas from the target. 

With just a single adaptive optics observation, it is still possible (though very unlikely) that an eclipsing binary could reside at the precise location of the star. This false positive scenario can be ruled out with multiple adaptive optics observations over a large temporal baseline. 
Over the nearly 17 years between the first deep archival adaptive optics observation of \name{} (as described in \citet{ammlervoneiff}) and our new observations reported in this paper, the star's apparent position has moved by over 250\,mas due to its proper motion. 
Both the archival adaptive optics observation taken in 2004 and our new Keck observation has a sensitivity to companions with contrast greater than 5.57 magnitudes at 250 mas separations (at near-infrared wavelengths, so even greater contrast at visible wavelengths where TESS observes). Therefore, combining these adaptive optics observations makes it possible to rule out all background eclipsing binaries bright enough to cause the transit signal of \pname{}. 

\item \textbf{Physically-Associated Eclipsing Binaries}: One final false positive scenario is that a physically associated (comoving) faint companion star to \name{} is an eclipsing binary. This cannot be completely ruled out with multiple adaptive optics observations over time, because the companion moves with the target and could remain so close to being blended even in the high-resolution observations. However, to remain undetected by our high-angular-resolution observations, an eclipsing binary companion would have to orbit at a distance less than about 3\,AU, and therefore with orbital periods less than about 5\,years. Such a stellar companion would be easily detected in the archival RV observations if it exists, except in extremely finely-tuned and unlikely (probability less than $10^{-4}$) face-on inclinations that would reduce the RV shifts to less than about 100\,$\mathrm{m\,s^{-1}}$. Therefore, with high confidence, we conclude that there are no physically associated eclipsing binaries that could mimic the signal of \pname{}. 
\end{itemize}

Since we can confidently rule out all of these possible false positive scenarios, we conclude that the overwhelmingly likeliest explanation for the observed transit signal is that \pname{} is a genuine exoplanet and consider it validated. %This is especially true since planet candidates in multi-planet systems are known to have lower \textit{a priori} false positive probabilities than solitary planet candidates \citep{latham, Lissauer2012, Rowe2014}, by up to about a factor of 50 for small planets detected by \TESS\ \citep{Guerrero2021}. 

\subsection{Statistical Validation with \texttt{TRICERATOPS}}
\label{sec:triceratops analysis}

\begin{figure}%[tbh!]
   \centering
   \includegraphics[width=0.48\textwidth]{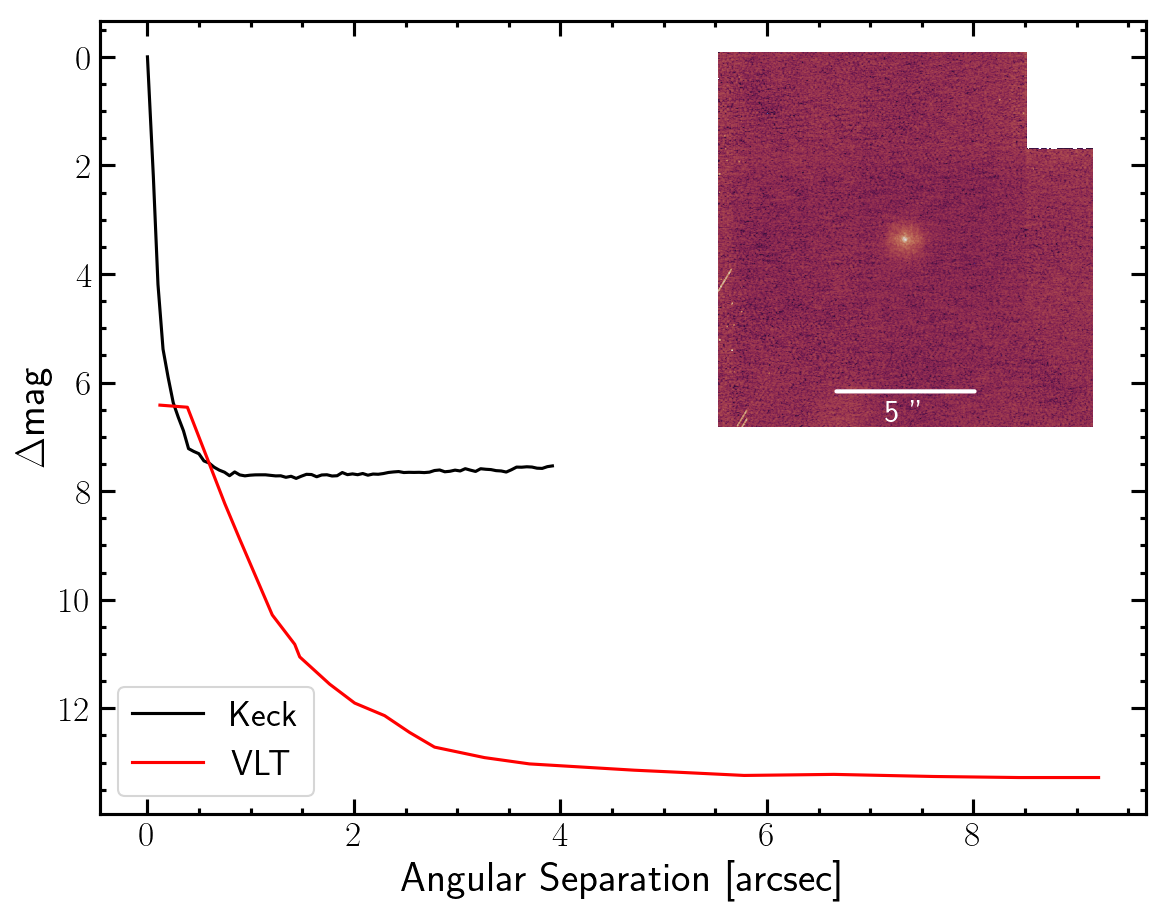}
   \caption{Contrast curves obtained from Keck/NIRC2 high-contrast observations (black line; Br-$\gamma$ filter) and VLT observations (red line; K filter). The 5$\sigma$ contrast limits for secondary sources are plotted against angular separation. 
   %AO contrast curves from the  imaging AO observations (black) and the VLT observations (red). We used the largest contrast at each separation from these two observations for our \texttt{TRICERATOPS} false positive analysis.
   }
   \label{fig:contrast}
\end{figure}

To double-check our conclusion that no false positive scenarios can mimic the signal of HD 63433 d, we employed the \texttt{TRICERATOPS} pipeline \citep{Giacalone2021} to calculate the likelihood of false positive scenarios. \texttt{TRICERATOPS} computes the probability that an input light curve originates from a wide range of scenarios using Bayesian inference. The pipeline takes into account several astrophysical false positive configurations, including those discussed above, and calculates the False Positive Probability (FPP), which is the probability that the observed transit is not caused by a planet (e.g., an eclipsing binary) and the Nearby False Positive Probability (NFPP), that is the probability that the signal comes from a nearby resolved star. A planet is considered statistically validated when the false positive scenarios can be ruled out with a certain level of confidence, typically the FPP$<0.01$ and the NFPP$<0.001$ \citep{Morton2012, Rowe2014, Morton2015, Giacalone2021}. We ran \texttt{TRICERATOPS} using the light curve for planet d, with the transit signals for planets b and c removed, phase-folded on the period from our \texttt{edmcmc} models. 
Furthermore, we also implemented the Keck NIRC2 high-contrast imaging AO observations described in Section \ref{sec:AO} as well as observations from the VLT reported in \cite{ammlervoneiff}, to better constrain the overall calculation.
The contrast curves corresponding to these two observations are shown in Figure~\ref{fig:contrast}.
%contrast plot reference here?
We obtained FPP$=(1.1\pm 2.5)\times 10^{-5}$ and NFPP$=0$ for the signal associated with \pname{}. 
This value for false positive probability falls well below the customary threshold of $10^{-3}$ for a statistically validated planet and agrees with the claims discussed in Section~\ref{sec:Vetting}. 

%\texttt{Triceratops} predicts that the most probable false-positive scenario is a background eclipsing binary (BEB) with a probability of $\sim 4\times 10^{-6}$.
%With this low probability in addition to our discussion in the previous section, we have ruled out the possibility of eclipsing binaries through archival and follow-up observations with high confidence. 

% 3.885134e-06 if we want to be more exact, but I don't have an uncertainty on this and I think the main point is just that it's very low

\section{Discussion}\label{sec:results}
\subsection{Physical Parameters of \pname{}}
Based on the results of our analysis, \pname{} can be described as a ``hot Earth.'' Our MCMC modeling, combined with derived stellar parameters, yields a radius of \prad, just slightly larger than the size of our home planet. The planet orbits \name{} every \porb{} (these errors are on the order of one second).
The planet has a semimajor axis that is about 20 times smaller than that of Earth (\semimajoraxis). 
We constrain the eccentricity to be moderate at most  ($< 0.52$ at 84\% confidence, and $< 0.73$ at 95\% confidence), but our posterior draws are consistent with a perfectly circular orbit. This would be in line with results from \textit{Kepler} for planets in multi-transiting systems \citep{vaneylenecc}.

{The orbital period ratios of planets b and d are close to resonance. However, since the ratio is slightly interior to the location of the resonance, it is unlikely that the system is in resonance, as the pile-up of resonant systems occurs slightly exterior to the resonance \citep{2012ApJ...750..114F}.}

The equilibrium temperature, $T_{eq}$, of \pname{} can be calculated assuming a circular orbit, an albedo ($\alpha$) of 0.3, and perfect heat redistribution, such that
\begin{equation}\label{noredistribution}
T_\mathrm{eq}=T_{\rm eff}(1 - \alpha)^{1/4}\sqrt{\frac{R_\star}{2a}},
\end{equation}
where $a$ is the planet's semi-major axis, $T_{\rm eff}$ is the stellar effective temperature, and $R_\star$ is the radius of the host star. 
Leveraging the values provided in this paper, this results in an equilibrium temperature of $T_\mathrm{eq} = 1040 \pm 40$\,K. 

However, Radial velocity observations \citep[e.g.,][]{rogers} have shown that planets the size of \pname{} almost always have predominantly rocky compositions and no thick atmospheres (although note that these observations come almost exclusively from stars older than \name{}, see Section~\ref{sec:massloss}). If \pname{} is indeed predominantly rocky, then it likely has no thick atmosphere that could efficiently redistribute heat, as assumed by Equation \ref{noredistribution}. Therefore, we estimate the dayside temperature of \pname{} under the assumption of a tidally locked blackbody (i.e., a rocky surface with no atmosphere). 
The assumption that the planet is tidally locked is supported by the tidal locking timescale, as calculated from \cite{Gladman1996}. More specifically, assuming a tidal $Q$ of 100 and an initial rotation rate of one cycle per day, we derive a tidal locking timescale of $t_{\mathrm{loc}}\approx10^3-10^4$\,yr,  which is much shorter than the age of the system.
According to \cite{2011ApJ...726...82C}, the apparent observed dayside temperature ($T_{\rm day}$) of a planet in a non-eccentric orbit, with no heat redistribution to its night side,\footnote{At a constant albedo, the planet's dayside temperature is greater than the equilibrium temperature by a factor of $\approx1.28$.} can be expressed as 
\begin{equation}
T_{\rm day} = [2/3(1-\alpha)]^{1/4} T_{\rm sub},\\
\end{equation}
where $T_{\rm sub} = T_{\rm eff}(R_\star/a)^{1/2}$ and represents the temperature at the planet's substellar point (the point on the planet's surface such that the host star is at the zenith). 
Assuming $\alpha = 0$ (similar to the measured albedo for the rocky planet Trappist-1b; \citealt{2023arXiv230314849G}), we calculate that the dayside temperature of \pname{} is approximately $T_{\rm day}$ = 1530\,K ($a = 0.0481$\,AU). 
The planetary temperature resembles other lava planets, such as CoRoT-7b and Kepler-10b (\citealt{Leger2009, Batalha2011}).
%CoRoT-7b: 1800-2600 K, Kepler-10b: 1833 K 

\subsection{\pname{} in Context}
\begin{figure}[t!]
    \centering
    \includegraphics[width=0.49\textwidth]{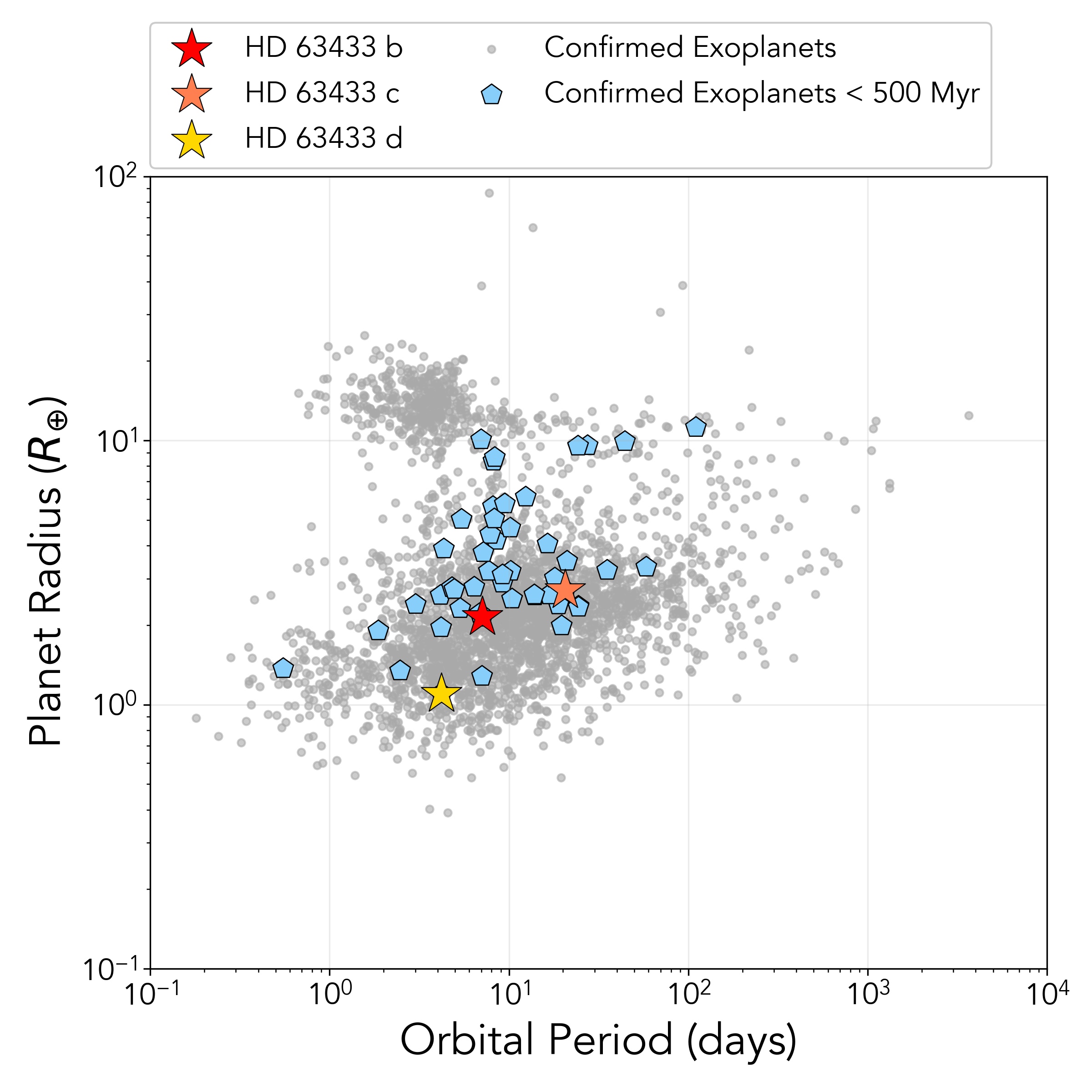}
    \caption{Planet radius versus orbital period for confirmed exoplanets (grey circles), confirmed exoplanets with reliable age estimates that are $<500$\,Myr (blue pentagons), and the three \name{} planets (b: red star, c: orange star, and d: yellow star). Note that there are few detections of young, Earth-sized planets. Those detected are in short-period orbits about their stellar hosts. For enhanced readability of the figure, we omit exoplanets detected via the direct imaging method, as their orbital separations are much larger.}
    \label{fig:periodvradius}
\end{figure}

HD 63433\,d's small size and precisely known young age, in combination with the proximity and brightness of its host star, make it stand out from the known population of exoplanets in several ways. 
Firstly, \pname{} is the smallest known planet with a precisely measured age that is less than 500\,Myr, making it an important touchstone for understanding how Earth-sized planets evolve. 
This is shown in Figure~\ref{fig:periodvradius}, which depicts the planet radius as a function of the orbital period for all confirmed exoplanets (NASA Exoplanet Archive, \citealt{akeson}).
While the full sample is shown in grey, we highlight in blue the relatively small subset of planets with precisely measured ages less than 500 Myr. 
For context, we considered a system age estimate to be precisely measured if its error estimate is at most 50\% of the estimated age value.
Note that the apparent lack of planets older than 500\,Myr with orbital periods greater than $10^{4}$\,days is because the technique used to find these objects (direct imaging) is biased toward young, bright systems.

\begin{figure}[htb!]
    %\centering
    \includegraphics[width=0.49\textwidth]{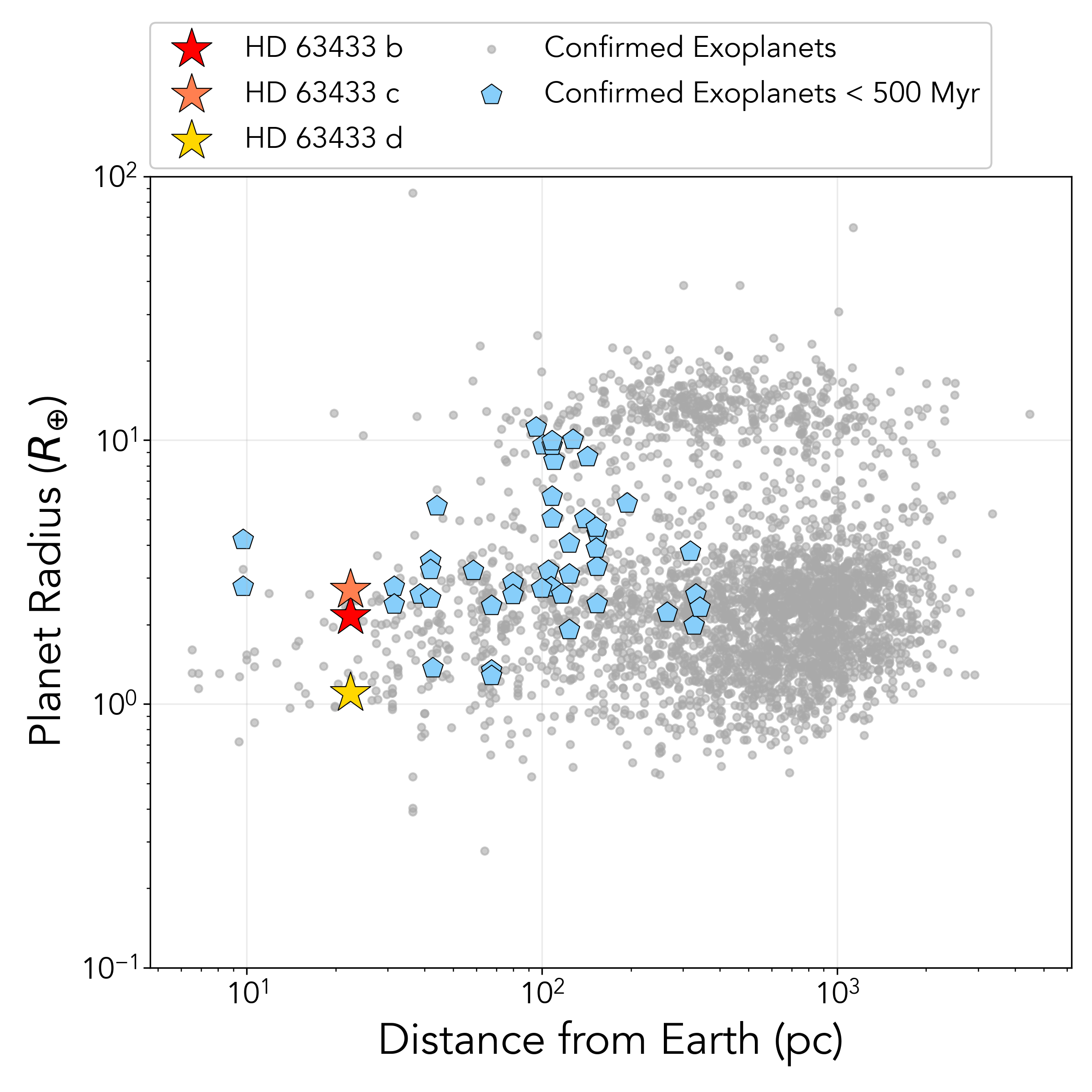}
    \caption{Distance from Earth versus planet radius for confirmed exoplanets (grey circles), confirmed exoplanets with reliable age estimates that are $<500$\,Myr (blue pentagons), and the three \name{} planets (b: red star; c: orange star; and d: yellow star). 
    Note that \pname{} is the nearest, young, Earth-sized exoplanet discovered to date.}
    \label{fig:radiusvdistance}
\end{figure}

\name{} is one of the nearest and brightest transiting planet hosts in the sky. According to the NASA Exoplanet Archive (accessed 2023 November 9), only seven transiting planet hosts\footnote{In order of ascending brightness: WASP 189, HR 858, HD 158259, 55 Cnc, pi Men, $\nu^2$ Lup, and HD 219134.} with brighter magnitudes in the \tess{} bandpass are known. None of these stars are particularly young or are members of clusters or associations.
Therefore, \name{} is the brightest known host of young transiting exoplanets in the sky. The high rate of stellar photons from \name{} striking \tess{}'s detectors enabled high enough light curve precision to detect the very shallow (100\,ppm) transits of this Earth-sized planet. 

HD 63433's brightness is due to its proximity (22\,pc) to the Solar system. Figure~\ref{fig:radiusvdistance} depicts the population of confirmed exoplanets as a function of planet radius and distance from Earth to the host star. Among young stars hosting transiting planets, only AU Mic lies closer to Earth than HD\,63433. HD 63433's close distance to Earth allows studies of atmospheric escape. \cite{zhang2022} searched for signatures of ongoing photoevaporation in the two previously known planets, and observed Lyman $\alpha$ absorption from HD~63433\,c, but not from HD~63433\,b. Using hydrodynamical simulations, they showed that if HD~63433\,b had a H/He atmosphere, its photoevaporative outflow would have been observable. Those results suggest that HD~63433\,b lost its primordial H/He atmosphere.
\pname{} experiences a stellar flux twice as large as that of HD~63433\,b; therefore, unless \pname{} is much more massive than standard mass-radius relations \citep[e.g.,][]{forecaster} predict, it likely lost its primordial H/He atmosphere too. Observations and models similar to those of \cite{zhang2022} could test this expectation and help understand the atmospheric evolution of the planet.
%the possibility of Lyman $\alpha$ spectroscopy to study atmospheric escape, which has already been performed for the two previously known planets by \citet{zhang2022}. Similar observations of \pname{} could be enlightening to understand whether any primordial H/He atmosphere remains. 

\subsection{\pname{}: A Case Study for Atmospheric Loss}
\label{sec:massloss}

Currently, one of the most important inquiries in exoplanet science is understanding in which circumstances planets keep or lose their thick primordial hydrogen/helium atmospheres, and what physical processes drive this phenomenon. Population-level studies have revealed the presence of a gap in the radius distribution of small planets \citep{owen, fulton, vaneylen, zeng}. Two of the leading mechanisms for understanding the formation of this feature, extreme ultraviolet photoevaporation \citep{owenphotoevap} and core-powered mass loss \citep{gupta} differ most significantly in timescales on which they act, so a promising way to distinguish these mechanisms is to study atmospheric escape and mass loss for planets with known ages. Recent studies \citep{Kreidberg2019,Crossfield2022,Greene2023,Zieba2023} have found that atmospheric loss is an important process for older terrestrial planets orbiting M-type stars. Given its radius well below the gap, HD 63433 d might be expected to have a rocky composition based on comparison with older planets, but if mass loss takes longer than $\approx$ 500\,Myr, it might still have a thick envelope. Because Earth-sized planets orbiting young, Sun-like stars have so far been difficult to detect, \pname{} presents a particularly compelling case study for atmospheric investigations of close-orbiting Earth-sized planets. 

\subsection{Future Work Characterizing HD 63433}

\pname{} presents intriguing and unique opportunities for future follow-up observations. It would be valuable to interrogate the planet's mass using precise radial velocities and determine whether the composition is indeed rocky, as expected based on observations of older planets. However, this will be challenging due to the high-amplitude stellar radial-velocity variability, which has limited attempts to measure the masses of the larger \name{} planets \citep{damasso, mallorquin}.   
Due to its proximity to Earth and high temperature, it may be feasible to utilize the James Webb Space Telescope (JWST) to detect the thermal emission from \pname{}. 
Several JWST MIRI programs have been designed to capture the thermal emission of rocky exoplanets, but  
\pname{} is younger than all terrestrial worlds with planned JWST emission observations. 
Although \pname{} is more luminous than many planets targeted for MIRI thermal emission observations, we estimate that the planet's eclipse depth in the mid-IR will only be approximately $20-30$\,ppm due to its orbit around a relatively large, sun-like star. MIRI should be capable of detecting such shallow eclipse depths; \cite{Bouwman2023} report 15\,ppm band-pass averaged precision in commissioning data, so the systematic noise floor should not be an issue. Moreover, the star's unusual brightness should provide plenty of photons to make these sensitive measurements. \cite{Kempton2018} presented an emission spectroscopy metric (ESM) for terrestrial planets that is proportional to the expected S/N of a JWST secondary eclipse detection at the center of the MIRI LRS band-pass, a wavelength of 7.5\,$\mu$m. For \pname{} we calculate an ESM of $7.1 \pm 0.3$, comparable to what the authors of that study found for planet GJ 1132b, which they used as their benchmark for a small rocky planet orbiting a nearby, bright M-dwarf. The precisely known age of this planet makes secondary eclipse observations particularly interesting for probing the presence of an atmosphere. 

% Adding Emission spectroscopy metric (ESM) of 7.1 +/- 0.3: 
% \citep{Kempton2018} - page 6 for description
% If we want to include the equation, but not sure we really need it: $$\mathrm{ESM}=4.29 \times 10^6 \times \frac{B_{7.5}\left(T_{\mathrm{day}}\right)}{B_{7.5}\left(T_*\right)} \times\left(\frac{R_p}{R_*}\right)^2 \times 10^{-m_K / 5}$$

Finally, \tess{} is scheduled to obtain another sector of observations of \name{} between 2023 November and 2023 December (Sector 72), and may continue observing the star in future mission extensions. These additional observations will improve the precision of parameters for the known planets and may reveal additional transit signals. 
In the \textit{Kepler} catalog, among the 277  systems with at least three transiting planet candidates, 35\% (98 total) host at least one additional planet.  {Although we find no additional transit signals in the BLS and TLS searches after removal of the three transit signatures, }the odds of \name{} itself hosting additional planets that may be revealed with additional observations are not insignificant. %\tbd{Future searches for transits would be performed best using the TLS algorithm due to its improved efficiency in shallow-transit detection over the BLS algorithm.}% fourth planetary parameters Additional \tess{} observations to detect more planets -- in Kepler, X \% of three-planet systems also have a 4th planet, so the odds are pretty good that if we keep looking and building SNR, we will find more planets. 

\subsection{Future Work Characterizing the Ursa Major Moving Group}
An important result of our work is that we have identified many new members of the Ursa Major moving group using new radial velocity measurements from \Gaia{}. We identified a set of 56 likely members based on kinematics and photometric rotation period analysis, which allowed us to significantly increase the number of Ursa Major sources with measured rotation periods and show a coherent rotation sequence with an age intermediate to that of the Pleiades and Praesepe clusters. The new observations from \Gaia{} promise significant further improvements to the membership and properties of the moving group. We excluded many stars from further analysis based on the lack of radial velocity measurements from \Gaia{}. Since \textit{Gaia} RVs are only measured with short exposures in the region of the Ca infrared triplet lines, hot stars with a few broad lines are likely systematically excluded. Ground-based follow-up may reveal that some of the sources we excluded, especially hot stars, indeed appear to be kinematic cluster members. Identifying new hot/massive stellar members could yield better constraints on the group's age through color-magnitude diagram fitting.
It may even be possible to measure asteroseismic oscillations for any members that exhibit pulsational signatures. Moreover, ground-based spectroscopic follow-up of potential UMa members to measure lithium abundances will provide another important constraint on the group's age. Future characterization of the Ursa Major group itself will ultimately enhance our knowledge of the \name{} multiplanet system. 

\section{Summary }\label{sec:summary}
In this work, we have presented the discovery and characterization of an Earth-sized planet orbiting the nearby, bright young star \name{}.
This is the third planet to be detected in the \name{} system.
The system is a likely member of the UMa moving group. We analyzed \tess{} data to study the planet candidate and analyzed the properties of the UMa moving group to assess its membership. 

\pname{} is an Earth-sized (\prad) planet orbiting a nearby (22\,pc), bright ($V\simeq6.9$\,mag), young (\age\,Myr) star that is reminiscent of a young Sun. 
\pname{} orbits its host star with a period of \porb{} interior to two other previously known mini Neptunes \citep{Mann2020}.
In addition, we identified a catalog of 56 likely UMa stellar members with kinematics matching the group's known motion and photometric rotation periods, consistent with its young estimated age. We re-establish membership of \name{} in the UMa group by showing that its color-magnitude diagram position, rotation period, lithium abundance, and kinematics are all highly consistent with these other sources. 

\pname{} is the closest planet to our Solar System with an Earth-like radius orbiting a young star.
Therefore, this is an appealing target for follow-up observations, offering an opportunity to reveal insights into the physics of exoplanet atmospheric mass loss. 
Between \pname{} and the two previously known larger planets, the \name{} system is poised to play an important role in our understanding of planetary system evolution in the first billion years after formation.

\acknowledgments
The authors would like to acknowledge helpful discussions with Richard Townsend,  Anne Noer Kolborg, and Lisa Kaltenegger.
MSF gratefully acknowledges the generous support provided by NASA through Hubble Fellowship grant HST-HF2-51493.001-A awarded by the Space Telescope Science Institute, which is operated by the Association of Universities for Research in Astronomy, In., for NASA, under the contract NAS 5-26555.
RY is grateful for support from a Doctoral Fellowship from the University of California Institute for Mexico and the United States (UCMEXUS) and CONACyT, a Texas Advanced Computing Center (TACC) Frontera Computational Science Fellowship, and a NASA FINESST award (21-ASTRO21-0068). AWM was supported by grants from the NSF CAREER program (AST-2143763) and NASA's exoplanet research program (XRP 80NSSC21K0393).

A portion of this work was performed at the Aspen Center for Physics, which is supported by National Science Foundation grant PHY-2210452.
This paper includes data collected by the \tess{} mission, which are publicly available from the Mikulski Archive for Space Telescopes (MAST). Funding for the \tess{} mission is provided by NASA’s Science Mission Directorate. 
This research has made use of the Exoplanet Follow-up Observation Program (ExoFOP; DOI: 10.26134/ExoFOP5) website, which is operated by the California Institute of Technology, under contract with the National Aeronautics and Space Administration under the Exoplanet Exploration Program.
This work has made use of data from the European Space Agency (ESA) mission \emph{Gaia},\footnote{\url{https://www.cosmos.esa.int/gaia}} processed by the \emph{Gaia} Data Processing and Analysis Consortium (DPAC)\footnote{\url{https://www.cosmos.esa.int/web/gaia/dpac/consortium}}. 
This research has made use of the VizieR catalogue access tool, CDS, Strasbourg, France. The original description of the VizieR service was published in A\&AS 143, 23. 
We acknowledge the use of public TOI Release data from pipelines at the \tess{} Science Office and the \tess{} Science Processing Operations Center. 
Resources supporting this work were provided by the NASA High-End Computing (HEC) Program through the NASA Advanced Supercomputing (NAS) Division at Ames Research Center for the production of the SPOC data products.
This publication makes use of data products from the Two Micron All Sky Survey, which is a joint project of the University of Massachusetts and the Infrared Processing and Analysis Center/California Institute of Technology, funded by the National Aeronautics and Space Administration and the National Science Foundation.
This work makes use of observations from the LCOGT network. Part of the LCOGT telescope time was granted by NOIRLab through the Mid-Scale Innovations Program (MSIP). MSIP is funded by NSF.
Funding for the TESS mission is provided by NASA's Science Mission Directorate. 
KAC and SNQ acknowledge support from the TESS mission via subaward s3449 from MIT. 
The research shown here acknowledges the use of the Hypatia Catalog Database, an online compilation of stellar abundance data as described in Hinkel et al. (2014, AJ, 148, 54), which was supported by NASA's Nexus for Exoplanet System Science (NExSS) research coordination network and the Vanderbilt Initiative in Data-Intensive Astrophysics (VIDA). ZLD would like to acknowledge the support of the MIT Presidential Fellowship and the MIT Collamore-Rogers Fellowship. ZLD  would like to acknowledge that this material is based upon work supported by the National Science Foundation Graduate Research Fellowship under Grant No. 1745302.

\facilities{\tess, \Gaia{} \citep{GaiaDR3}, Mikulski Archive for Space Telescopes \citep{MAST},  ROentgen SATellite (\emph{ROSAT}), Wide-field Infrared Survey Explorer (\emph{WISE}), LCOGT 1m (Sinistro), LCOGT 1m (NRES), SMARTS 1.5m (CHIRON), Tillinghast 1.5m (TRES), SOAR (Goodman), TNG (HARPS-N), OHP 1.93m (ELODIE), OHP 1.93m (SOPHIE), Shane 3m (Hamilton)}

\software{\texttt{AstroImageJ} \citep{Collins2017}, \texttt{astroquery} \citep{astroquery}, BANZAI \citep{banzai}, \texttt{batman} \citep{batman}, \texttt{corner.py} \citep{corner}, \texttt{edmcmc} \citep{vanderburgedmcmc}, \texttt{Lightkurve} \citep{lightkurve}, matplotlib \citep{matplotlib}, PAdova and TRieste Stellar Evolution Code \citep{parsec2012}, \texttt{PyAstronomy} \citep{pya}, \texttt{starrotate} \citep{starrotate}, \texttt{TAPIR} \citep{Jensen:2013}, \texttt{TESSCUT} \citep{tesscut}}
\vspace{5mm}
\bibliographystyle{aasjournalmod}
\bibliography{bibliography.bib}

\begin{deluxetable*}{lccc|cccccc}
\tabletypesize{\scriptsize}
\tablewidth{0pt}
\tablecaption{Transit-Fit Parameters. \label{tab:transfit} }
\tablehead{\colhead{Parameter} & \colhead{planet\,b} & \colhead{planet\,c} & \colhead{planet\,d} & \colhead{planet\,b} & \colhead{planet\,c} & \colhead{planet\,d} \\ 
\colhead{} & \multicolumn{3}{c}{$e$,$\omega$ fixed} &\multicolumn{3}{c}{$e$,$\omega$ free}}
\startdata
\multicolumn{7}{c}{Transit Fit Parameters} \\
\hline
$T_0$ (TJD) & $1916.45286\pm 0.00042$ & $1844.05971^{ +0.00052}_{-0.00051}$   &
$2373.82337^{ +0.00104}_{-0.00073}$ 
& $1916.45262^{ +0.00062}_{-0.00065}$ & $1844.05971^{ +0.00060}_{-0.00057}$ & $2373.82345^{ +0.00113}_{-0.00083}$ \\
$P$ (days) & $7.1079342\pm 0.0000049$ & $20.543784\pm 0.000016$   
& $4.209078^{ +0.000016}_{-0.000022}$ &
$7.1079379\pm0.0000054$ & $20.543791^{ +0.000017}_{-0.000018}$
& $4.209075^{ +0.000012}_{-0.000023}$ \\
$R_P/R_{\star}$ & $0.02116\pm 0.00011$ & $0.02578^{ +0.00014}_{-0.00013}$   
& $0.01090\pm 0.00016$
& $0.02123^{ +0.00050}_{-0.00034}$ & $0.02534^{ +0.00033}_{-0.00026}$ &
$0.01079^{ +0.00022}_{-0.00018}$ \\
$\rho_{\star}$ ($\rho_{\odot}$) & \multicolumn{3}{c}{ $1.11^{ +0.18}_{-0.21}$  } & \multicolumn{3}{c}{ $1.27 \pm 0.19$  } \\
$q_{1,1}$ & \multicolumn{3}{c}{ $0.150^{ +0.027}_{-0.024}$  } & \multicolumn{3}{c}{ $0.212^{ +0.029}_{-0.026}$  } \\
$q_{2,1}$ & \multicolumn{3}{c}{$0.0046^{ +0.0076}_{-0.0034}$ } & \multicolumn{3}{c}{ $0.382^{ +0.048}_{-0.047}$ } \\
$\sqrt{e}\sin\omega$ & 0 (fixed) & 0 (fixed) & 0 (fixed) & $0.29^{ +0.18}_{-0.25}$ & $-0.08^{ +0.26}_{-0.21}$ & $0.14^{ +0.23}_{-0.19}$\\
$\sqrt{e}\cos\omega$ & 0 (fixed) & 0 (fixed) & 0 (fixed) & 
$0.0023^{ +0.4767}_{-0.4680}$ & $-0.0084^{ +0.5822}_{-0.5520}$ & $-0.0038^{ +0.4865}_{-0.4709}$ \\
$i$ ($^{\circ}$) & $88.71^{ +0.49}_{-0.44}$ & $88.98^{ +0.11}_{-0.13}$ & $88.26^{ +0.83}_{-0.66}$   
& $88.49^{ +0.87}_{-0.40}$ & $89.28^{ +0.40}_{-0.22}$ 
& $88.73^{ +0.85}_{-1.06}$  \\
\hline
\multicolumn{7}{c}{Derived Parameters}\\
\hline
$a/R_{\star}$ & $16.08^{ +0.58}_{-0.72}$ & $32.6^{ +1.2}_{-1.5}$ & $11.34^{ +0.41}_{-0.51}$ &
$16.83^{ +0.57}_{-0.63}$ & $34.15^{ +1.16}_{-1.29}$
& $11.87^{ +0.40}_{-0.45}$ \\
$b$ & $0.31^{ +0.11}_{-0.13}$ & $0.56\pm 0.05$
& $0.31^{ +0.11}_{-0.16}$ 
& $0.38^{ +0.09}_{-0.21}$ & $0.37^{ +0.14}_{-0.21}$
& $0.23 \pm 0.16$ \\
$T_{14}$ (days) & $0.13422^{ +0.00045}_{-0.00041}$ & $0.16883^{ +0.00073}_{-0.00061}$ & 
$0.1118^{ +0.0013}_{-0.0022}$&
$0.122^{ +0.012}_{-0.012}$  & $0.171^{ +0.013}_{-0.014}$ &
$0.1077^{ +0.0056}_{-0.0149}$ \\
$T_{23}$ (days) & $0.12801^{ +0.00042}_{-0.00047}$ & $0.15655^{ +0.00060}_{-0.00073}$ & 
$0.1091^{ +0.0013}_{-0.0023}$ &
$0.115 \pm 0.012$  & $0.161^{ +0.014}_{-0.013}$ &
$0.105^{ +0.006}_{-0.015}$ \\
$e$ & 0 (fixed) & 0 (fixed) & 0 (fixed) & $0.24^{ +0.27}_{-0.18}$   & $0.21^{ +0.33}_{-0.14}$ & 
$0.16^{ +0.36}_{-0.12}$\\
$R_p$ ($R_\oplus$)$^1$ & $2.105\pm 0.079$ & $2.565\pm 0.097$
& $1.084\pm 0.043$ 
& $2.112^{ +0.093}_{-0.086}$ & $2.521\pm0.1$
& $1.073^{ +0.046}_{-0.044}$\\
$a$ (AU) & $0.0682^{ +0.0035}_{-0.0040}$ & $0.1382^{ +0.0072}_{-0.0082}$
& $0.0481^{ +0.0025}_{-0.0028}$ 
& $0.0714^{ +0.0036}_{-0.0038}$  & $0.1448^{ +0.0073}_{-0.0077}$
&$0.0503^{ +0.0025}_{-0.0027}$\\
\enddata
\tablenotetext{1}{$R_p$ derived using the $R_*$ value from Table~\ref{tab:prop}}
\end{deluxetable*}

\begin{deluxetable*}{rrrrrrrrrrr}
\centering
\tabletypesize{\scriptsize}
\tablewidth{0pt}
\centering
\tablecaption{Properties of the 56 likely members of UMaG included in our catalog (listed in ascending order by reddened-corrected color, \texttt{BP-RP\_abs}). While the rotation periods and errors (\texttt{Prot} and \texttt{e\_Prot}) were determined by this work, all other properties were obtained from \Gaia{}. We show the properties of \name{} in the top, bold row. Comments about individual catalog sources are not included here, but are available in the digital version of the catalog. \label{tab:UMaG_rot}}
\tablehead{\colhead{TICID} & \colhead{RA} & \colhead{DEC} & \colhead{Gmag} & \colhead{Gmag\_abs$^{1}$} & \colhead{BP-RP} & \colhead{BP-RP\_abs$^{2}$} & \colhead{Dist} & \colhead{RUWE} & \colhead{Prot} & \colhead{e\_Prot} \\
  & \colhead{(deg)} & \colhead{(deg)} & \colhead{(mag)} & \colhead{(mag)} & \colhead{(mag)} & \colhead{(mag)} & \colhead{(pc)} &  & \colhead{(d)} & \colhead{(d)} \\}
\startdata
\textbf{130181866} & \textbf{117.479} & \textbf{27.363} & \textbf{6.737} & \textbf{4.988} & \textbf{0.854} & \textbf{0.854} & \textbf{22.34} & \textbf{0.99} & \textbf{6.4} & \textbf{0.6}\\
\hline
235682463 & 287.2912 & 76.5605 & 5.032 & 2.846 & 0.445 & 0.445 & 27.46 & 1.10 & 0.77 & 0.1\\
86433449 & 214.8178 & 13.0043 & 5.289 & 3.205 & 0.549 & 0.549 & 26.09 & 1.10 & 1.5 & 0.2\\
408908804 & 92.0267 & -15.9012 & 6.843 & 3.516 & 0.597 & 0.597 & 46.23 & 0.93 & 1.7 & 0.2\\
362747897 & 247.1172 & -70.0844 & 4.744 & 4.336 & 0.737 & 0.737 & 12.04 & 2.11 & 8.2 & 0.8\\
283792884 & 63.8700 & 6.1869 & 6.175 & 4.456 & 0.743 & 0.743 & 22.05 & 1.10 & 6.5 & 0.7\\
157272202 & 198.4042 & 56.7083 & 6.663 & 4.659 & 0.780 & 0.780 & 25.12 & 1.00 & 6.0 & 0.6\\
329574145 & 271.5988 & -36.0198 & 5.805 & 4.638 & 0.781 & 0.781 & 17.10 & 0.55 & 6.0 & 0.6\\
417762326 & 129.7988 & 65.0209 & 5.499 & 4.701 & 0.804 & 0.804 & 14.45 & 1.17 & 4.9 & 0.5\\
 24910401 & 27.3473 & -10.7036 & 6.590 & 4.695 & 0.808 & 0.808 & 23.21 & 7.40 & 8.7 & 0.9 \\
283792891 & 63.8574 & 6.1997 & 6.778 & 5.059 & 0.860 & 0.860 & 22.05 & 1.05 & 6.5 & 0.7\\
 88659764 & 152.5783 & 38.4135 & 8.685 & 5.013 & 0.864 & 0.863 & 53.89 & 1.08 & 7.3 & 0.7\\
117881543 & 96.8361 & -33.1140 & 8.228 & 5.002 & 0.865 & 0.865 & 44.12 & 1.12 & 7.3 & 0.7\\
366351891 & 125.7439 & 7.6304 & 8.990 & 5.458 & 0.954 & 0.954 & 50.77 & 0.89 & 5.2 & 0.5\\
366351890 & 125.7473 & 7.6309 & 9.031 & 5.495 & 0.959 & 0.957 & 50.41 & 0.93 & 5.2 & 0.5\\
415563103 & 91.6687 & 15.5421 & 6.539 & 5.550 & 0.984 & 0.984 & 15.76 & 0.98 & 8.1 & 0.8\\
176471832 & 86.9683 & -2.7606 & 9.104 & 5.693 & 1.015 & 1.009 & 47.45 & 0.97 & 8.2 & 0.8\\
16045498 & 113.2524 & 37.0298 & 7.426 & 5.848 & 1.057 & 1.057 & 20.65 & 1.03 & 8.0 & 0.8\\
149852612 & 44.6862 & 40.0723 & 9.559 & 6.152 & 1.081 & 1.081 & 47.94 & 1.21 & 9.5 & 1.0\\
308056339 & 99.4094 & 16.2704 & 9.670 & 6.064 & 1.104 & 1.104 & 52.54 & 1.17 & 15.2 & 1.5\\
288158059 & 98.2995 & -17.4905 & 8.909 & 6.126 & 1.120 & 1.119 & 36.05 & 1.11 & 9.0 & 1.0\\
99381773 & 190.4355 & 55.7247 & 7.985 & 6.203 & 1.156 & 1.156 & 22.69 & 0.97 & 12.0 & 1.2\\
224305606 & 188.9637 & 51.2215 & 8.258 & 6.167 & 1.169 & 1.169 & 26.18 & 1.31 & 9.0 & 1.0 \\
105577896 & 153.3202 & 42.3554 & 10.494 & 6.710 & 1.298 & 1.298 & 56.98 & 1.09 & 9.6 & 1.0\\
198381449 & 257.5438 & 54.4944 & 8.475 & 6.839 & 1.397 & 1.397 & 21.23 & 0.97 & 11.6 & 1.2\\
419366667 & 121.0079 & -35.2761 & 10.388 & 7.077 & 1.489 & 1.444 & 44.73 & 2.89 & 11.7 & 1.2\\
198381445 & 257.5515 & 54.4901 & 8.851 & 7.214 & 1.534 & 1.534 & 21.21 & 0.97 & 11.6 & 1.2\\
 97526849 & 167.4684 & 21.6255 & 10.508 & 7.468 & 1.634 & 1.634 & 40.46 & 1.04 & 12.7 & 1.3\\
 95618424 & 121.4938 & 26.2814 & 9.475 & 8.158 & 1.903 & 1.902 & 18.34 & 1.14 & 13.6 & 1.4\\
176469654 & 86.8249 & -0.0136 & 10.207 & 8.134 & 1.914 & 1.912 & 25.83 & 1.05 & 15.0 & 1.5\\
 97488127 & 165.6598 & 21.9671 & 8.811 & 8.426 & 2.006 & 2.005 & 11.92 & 1.12 & 13.6 & 1.4\\
464452462 & 282.6110 & -62.0510 & 9.883 & 8.765 & 2.087 & 2.087 & 16.76 & 0.97 & 8.0 & 0.8\\
159503663 & 246.4935 & 83.4067 & 11.028 & 8.722 & 2.121 & 2.119 & 28.88 & 1.16 & 16.7 & 1.7\\
139053298 & 182.2063 & 30.3503 & 11.481 & 9.223 & 2.272 & 2.272 & 28.13 & 1.05 & 8.6 & 0.9\\
386955374 & 172.8537 & 13.7454 & 11.844 & 9.049 & 2.289 & 2.289 & 36.27 & 0.97 & 4.5 & 0.5\\
390733970 & 44.0827 & 61.6896 & 12.289 & 9.592 & 2.369 & 2.369 & 34.72 & 1.13 & 4.5 & 0.2\\
290970527 & 309.0346 & -36.1199 & 10.587 & 9.449 & 2.395 & 2.395 & 16.89 & 1.17 & 2.1 & 0.2\\
165599288 & 227.2838 & 59.0738 & 11.884 & 9.545 & 2.439 & 2.437 & 29.18 & 1.34 & 4.0 & 0.4\\
 98506634 & 227.6127 & 13.4671 & 12.545 & 10.260 & 2.642 & 2.642 & 28.58 & 1.30 & 1.8 & 0.2\\
 77896213 & 78.8634 & -35.8070 & 13.062 & 10.437 & 2.649 & 2.646 & 33.33 & 1.23 & 10.6 & 1.0\\
20118579 & 130.1653 & 45.0045 & 13.578 & 10.418 & 2.709 & 2.709 & 42.79 & 1.26 & 7.3 & 0.7\\
313110473 & 211.0967 & 72.4130 & 13.115 & 10.452 & 2.740 & 2.740 & 33.99 & 1.27 & 4.9 & 0.5\\
290474796 & 139.0424 & 1.8858 & 11.689 & 10.716 & 2.764 & 2.764 & 15.64 & 1.12 & 1.5 & 0.2\\
309661100 & 116.8078 & 50.3440 & 11.782 & 11.035 & 2.799 & 2.799 & 14.11 & 1.41 & 0.77 & 0.1\\
219463771 & 208.4116 & 77.6189 & 11.540 & 10.929 & 2.803 & 2.803 & 13.25 & 1.20 & 1.25 & 0.1\\
291962446 & 106.2752 & 4.3343 & 14.581 & 11.144 & 2.825 & 2.824 & 48.60 & 1.08 & 0.75 & 0.8\\
80949873 & 122.0214 & 59.6077 & 13.498 & 10.916 & 2.839 & 2.834 & 32.59 & 1.23 & 2.7 & 0.3\\
72882364 & 86.1131 & -4.1004 & 13.247 & 11.365 & 2.866 & 2.866 & 23.82 & 1.17 & 0.9 & 0.1\\
72636078 & 218.0450 & 16.0137 & 12.241 & 11.251 & 2.867 & 2.867 & 15.71 & 1.36 & 0.77 & 0.1\\
162131472 & 249.4300 & -20.2261 & 12.326 & 10.856 & 2.872 & 2.872 & 19.68 & 1.30 & 3.4 & 0.4\\
61627029 & 78.3220 & 27.6173 & 13.765 & 11.126 & 2.941 & 2.898 & 32.70 & 1.12 & 1.4 & 0.1\\
219787770 & 264.5203 & 65.1575 & 13.575 & 10.764 & 2.942 & 2.898 & 35.44 & 1.29 & 0.4 & 0.1\\
396948416 & 179.6839 & 10.3325 & 14.431 & 11.299 & 2.955 & 2.953 & 42.23 & 1.23 & 0.9 & 0.1\\
21220201 & 128.9103 & 16.4825 & 14.346 & 11.744 & 2.983 & 2.983 & 32.92 & 1.16 & 0.2 & 0.1\\
17992446 & 171.0182 & 38.1364 & 13.008 & 11.677 & 3.017 & 3.017 & 18.43 & 1.41 & 0.5 & 0.1\\
141821609 & 206.2412 & 55.4728 & 13.930 & 11.846 & 3.025 & 3.025 & 26.02 & 1.33 & 0.3 & 0.1
\enddata
\tablenotetext{1}{The absolute Gaia magnitude, such that $G_\mathrm{abs}=G+5\times \log_{10}(\varpi/100)-A_{\mathrm{G}}.$}
\tablenotetext{2}{The absolute BP-RP color, such that BP-RP\_abs=BP-RP-E(BP-RP).}
\end{deluxetable*}

\end{document}